\documentclass{mn2e}
\usepackage{epsfig,natbib2,natbibmnfix}
\usepackage{amssymb}
\usepackage{amsbsy}
\usepackage{multirow}
\usepackage{multicol}
\usepackage[varg]{txfonts}
\usepackage{color}
\usepackage{mathrsfs}
\usepackage{astrojournals}
\usepackage{subfigure}
\newcommand{\fref}[1]{Fig.~\ref{#1}}

\newcommand{\eref}[1]{equation~\ref{#1}}
\newcommand{\cref}[1]{Chapter~\ref{#1}}

\def \mathbi#1{\textbf{\em #1}}

\begin{document}
%%%%%%%%%%%%%%%%%%%%%%%%%%%%%%%%%%%%%%%%%%%%%%%%%%%%%%%%%%%%%%%%%%%%%%%%%%%%%%
%% Title Details and Page Header                                            %%
%%%%%%%%%%%%%%%%%%%%%%%%%%%%%%%%%%%%%%%%%%%%%%%%%%%%%%%%%%%%%%%%%%%%%%%%%%%%%%
\title[Broken discs]{Broken discs: warp propagation in accretion
  discs}

\author[C.~J.~Nixon \& A.~R.~King] 
{
\parbox{5in}{Christopher~J. Nixon$^1$ and Andrew~R. King$^1$}
\vspace{0.1in} 
 \\ $^1$ Department of Physics \& Astronomy, University of Leicester,
 Leicester LE1 7RH UK  
}

\maketitle

%%%%%%%%%%%%%%%%%%%%%%%%%%%%%%%%%%%%%%%%%%%%%%%%%%%%%%%%%%%%%%%%%%%%%%%%%%%%%%
%% Abstract, Keywords and contact details                                   %%
%%%%%%%%%%%%%%%%%%%%%%%%%%%%%%%%%%%%%%%%%%%%%%%%%%%%%%%%%%%%%%%%%%%%%%%%%%%%%%
\begin{abstract}
We simulate the viscous evolution of an accretion disc around a spinning black
hole. In general any such disc is misaligned, and warped by the
Lense--Thirring effect. Unlike previous studies we use effective viscosities
constrained to be consistent with the internal fluid dynamics of the disc. We
find that nonlinear fluid effects, which reduce the effective viscosities in
warped regions, can promote the breaking of the disc into two distinct
planes. This occurs when the Shakura \& Sunyaev dimensionless viscosity
parameter $\alpha$ is $\lesssim 0.3$ and the initial angle of misalignment
between the disc and hole is $\gtrsim 45^{\circ}$. The break can be a
long--lived feature, propagating outwards in the disc on the usual alignment
timescale, after which the disc is fully co-- or counter--aligned with the
hole. Such a break in the disc may be significant in systems where we know
the inclination of the outer accretion disc to the line of sight, such as some
X--ray binaries: the inner disc, and so any jets, may be noticeably misaligned
with respect to the orbital plane.
\end{abstract}


\footnotetext[1]{E--mail: chris.nixon@astro.le.ac.uk}

%%%%%%%%%%%%%%%%%%%%%%%%%%%%%%%%%%%%%%%%%%%%%%%%%%%%%%%%%%%%%%%%%%%%%%%%%%%%%%
%% Introduction                                                             %%
%%%%%%%%%%%%%%%%%%%%%%%%%%%%%%%%%%%%%%%%%%%%%%%%%%%%%%%%%%%%%%%%%%%%%%%%%%%%%%
\section{Introduction}
\label{intro}
In realistic astrophysical situations the angular momentum of a spinning black
hole may be significantly misaligned with respect to a surrounding accretion
disc. This is expected in black hole X--ray binaries where the hole may have
received a significant kick during formation (\citealt{Shklovskii1970};
\citealt{Sutantyo1978}; \citealt*{Arzoumanianetal2002};
\citealt{Hobbsetal2005}). It is also likely for accretion on to a supermassive
black hole (SMBH) in the centre of a galaxy. The scale of the SMBH is so small
compared with the galaxy that any process which drives gas down to the
galactic centre is unlikely to have any special orientation with respect to
the SMBH spin. The spin direction is set by the SMBH's accretion history and
not by whatever causes the next accretion event. Processes such as star
formation, merging of satellite galaxies, gas cloud collisions and other
galactic phenomena likely to drive gas inwards are chaotic, so it is
reasonable to expect a random spread of angles between the angular momentum of
infalling gas and that of the hole (cf \citealt{KP2006,KP2007};
\citealt*{Kingetal2008}). This gas is likely to form a ring around the SMBH
which then viscously spreads into a misaligned accretion disc.

The Lense--Thirring (hereafter LT) effect of a spinning black hole
causes non--equatorial orbits to precess at rates dependent on the
distance from the hole \citep{LT1918}. In an accretion disc, this
differential precession causes a warp which propagates through the
disc. This is the Bardeen--Petterson effect \citep{BP1975}. In this
paper we restrict our attention to viscous Keplerian accretion discs
with negligible self--gravity.  \citet{PP1983} linearised the
equations of fluid dynamics to derive the first consistent equation
governing the evolution of a warped disc of this type, assuming a tilt
angle smaller than the disc angular semi--thickness. They introduced
two effective kinematic viscosities controlling the angular momentum
transport in the disc. In principle these two quantities depend on the
nature of the warp: $\nu_{1}$ governs the usual radial communication
of the component of angular momentum perpendicular to the plane of the
disc due to differential rotation (for a flat disc this is the Shakura
\& Sunyaev alpha viscosity), while $\nu_{2}$ governs the radial
communication of the component of angular momentum parallel to the
local orbital plane; this acts to flatten tilted rings of gas.
\citet{PP1983} also showed that conservation laws require a relation
between $\nu_{1}$ and $\nu_{2}$. For small warps this is
\begin{equation}
\nu_{2} = \frac{\nu_{1}}{2\alpha^{2}},
\label{nu12}
\end{equation}  
where $\alpha$ is the usual dimensionless viscosity parameter for discs
\citep{SS1973} assumed $\ll 1$ by \citet{PP1983}. This gives
\begin{equation}
\alpha_{2} = \frac{1}{2\alpha},
\label{alpha12}
\end{equation}      
(again for small $\alpha$) where $\alpha_{2}$ is an equivalent
dimensionless viscosity parameter for $\nu_{2}$.

\citet{Pringle1992} derived an evolution equation valid for larger
warps expressing global angular momentum conservation for a disc
composed of rings of gas with no internal degrees of
freedom. \citet{Pringle1992} went on to develop a numerical technique
for integrating this equation over rings of gas to study the viscous
evolution of a time--dependent warped disc. \citet{LP2006} used this
to study the Bardeen--Petterson effect. They assumed constant
effective viscosities, related as expected for small amplitude warps
(eq.~\ref{nu12}). This treatment suggests that a disc warp must always
straighten itself out on a timescale shorter than the accretion
timescale (assuming only the LT torque and the internal disc torques,
i.e. no other external torques), because (\ref{nu12}) implies $\nu_{2}
\gg \nu_{1}$ for typical values of $\alpha$. The realignment process
enhances the accretion rate by inducing extra dissipation
\citep{LP2006}.

The analysis of \citet{Pringle1992} did not allow for any internal
degrees of freedom within the gas rings, and was described as the
`naive approach' in \citet{PP1983}. However \citet{Ogilvie1999}
derived equations of motion directly from the full three--dimensional
fluid dynamical equations with an assumed isotropic viscosity. This
confirmed that the equations of \citet{Pringle1992} are formally valid
only when two differences are taken into account (or can be safely
neglected). The first is that the \citet{Pringle1992} equations omit a
torque causing rings to precess if they are tilted with respect to
their neighbours. In a viscous Keplerian disc where $\alpha \ll 1$ but
not close to zero, this torque is much smaller than the usual viscous
torques included by \citet{Pringle1992} and so can be neglected in
time--dependent calculations.  (however for completeness we include
the torque in this work).  Our main focus, however, is on the second
extension. The quantities $\nu_{1}$ and $\nu_{2}$ are functions of the
disc structure and so depend on the warp amplitude $\left|\psi\right|
= R \left|{\partial} \boldsymbol{\ell}/{\partial} R \right|$, where
$\boldsymbol{\ell}$ is the unit angular momentum vector and $R$ is the
spherical radius coordinate. \citet{Ogilvie1999} used the equations of
fluid dynamics to determine the relation between the effective
viscosities for a general warp amplitude. In contrast to an assumption
of constant effective viscosities, this analysis suggests that these
quantities {\it drop} in a warped region, making it much harder for
the disc to straighten any twisted rings of gas. Indeed, if the warp
becomes large enough the disc may break into distinct planes with only
a tenuous connection. We expect this to occur in large warps, as
nonlinear corrections to the fluid flow are significant for warp
amplitudes $\left|\psi\right| \gtrsim \alpha$ \citep{Ogilvie1999}.

There have been several theoretical investigations which show evidence of
breaking discs. \citet{LP1997} report an example in Smoothed Particle
Hydrodynamics (SPH) simulations of a circumbinary disc with a strong radial
density decrease. This is suggestive, since the evolution of a misaligned
circumbinary disc is qualitatively similar to that of a misaligned disc around
a spinning black hole \citep*{Nixonetal2011b}. SPH simulations by
\citet{LP2010} also found disc breaking. They assumed the presence of a
large--amplitude warp and followed its viscous development. Their results
agreed with those expected for the constrained viscosities derived in
\citet{Ogilvie1999}, and showed that the disc could break if the warp
amplitude was large enough (note that this paper did not attempt to show that
the LT precession could {\it produce} such a warp).

In this paper we explore disc evolution under the LT effect with general warp
amplitudes, and the effective viscosities constrained as in
\citet{Ogilvie1999}. We explore the evolution for a range of $\alpha$ values
and different degrees of misalignment between the disc and hole. We aim to
discover if the internal torques in an accretion disc can sustain a smooth
warped configuration under the LT torque, or whether instead the disc breaks
into distinct planes.

%%%%%%%%%%%%%%%%%%%%%%%%%%%%%%%%%%%%%%%%%%%%%%%%%%%%%%%%%%%%%%%%%%%%%%%%%%%%%%
%% Bullet points                                                            %%
%%%%%%%%%%%%%%%%%%%%%%%%%%%%%%%%%%%%%%%%%%%%%%%%%%%%%%%%%%%%%%%%%%%%%%%%%%%%%%
\section{Numerical method}
\label{code}

\begin{figure*}
  \centering
    \subfigure{\includegraphics[angle=0,width=0.33\textwidth]
       {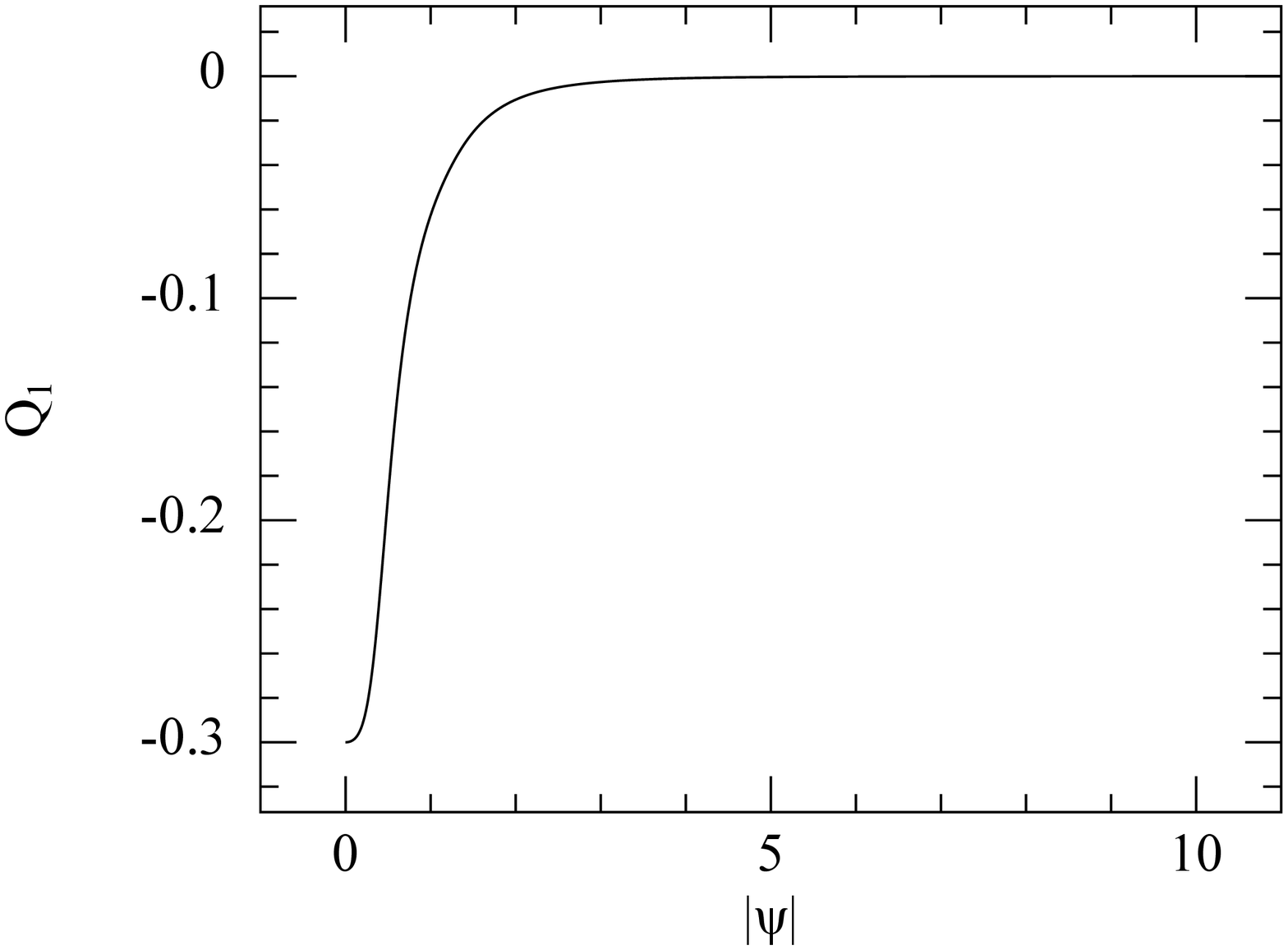}}
    \subfigure{\includegraphics[angle=0,width=0.33\textwidth]
       {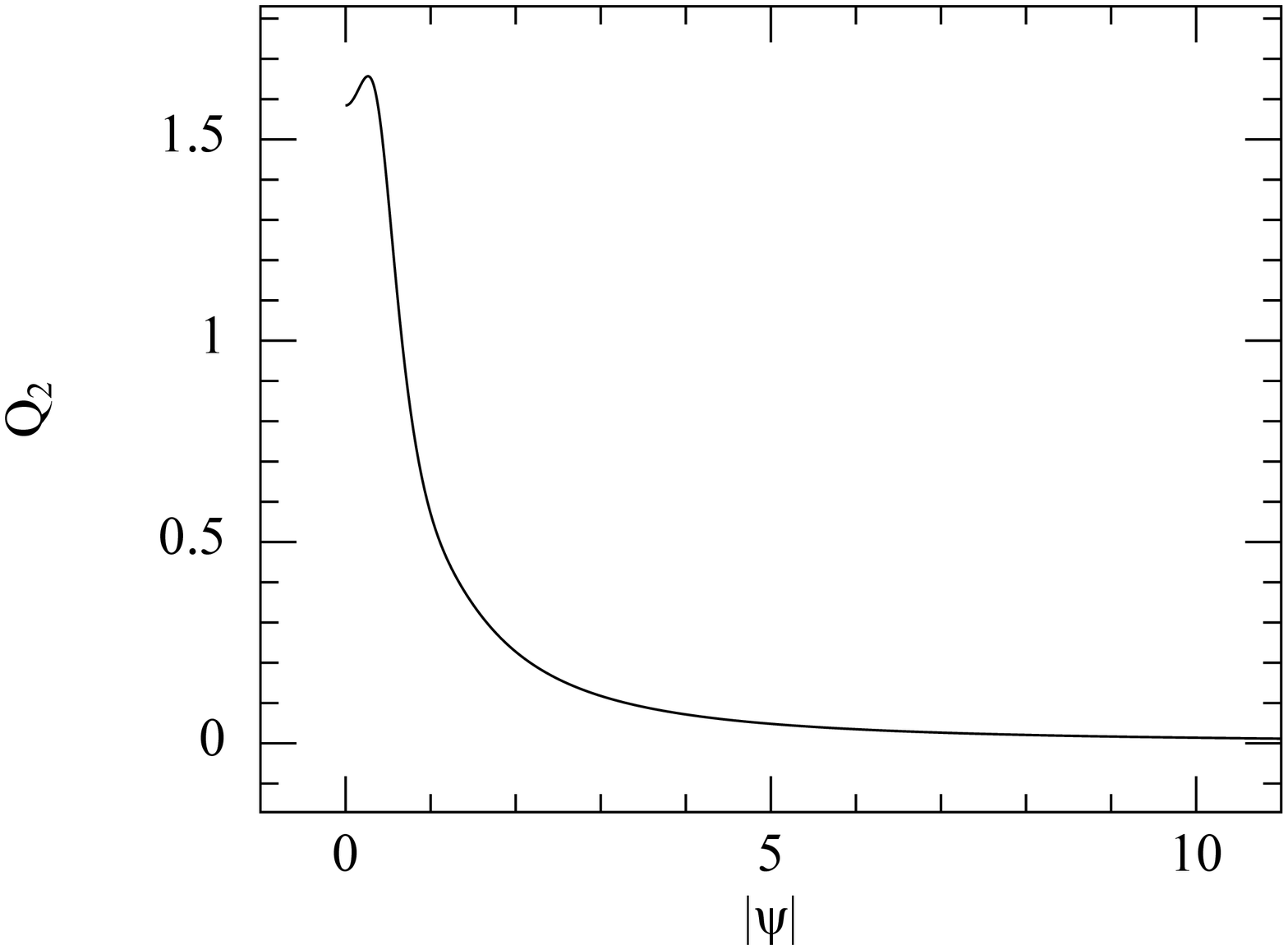}}
    \subfigure{\includegraphics[angle=0,width=0.33\textwidth]
       {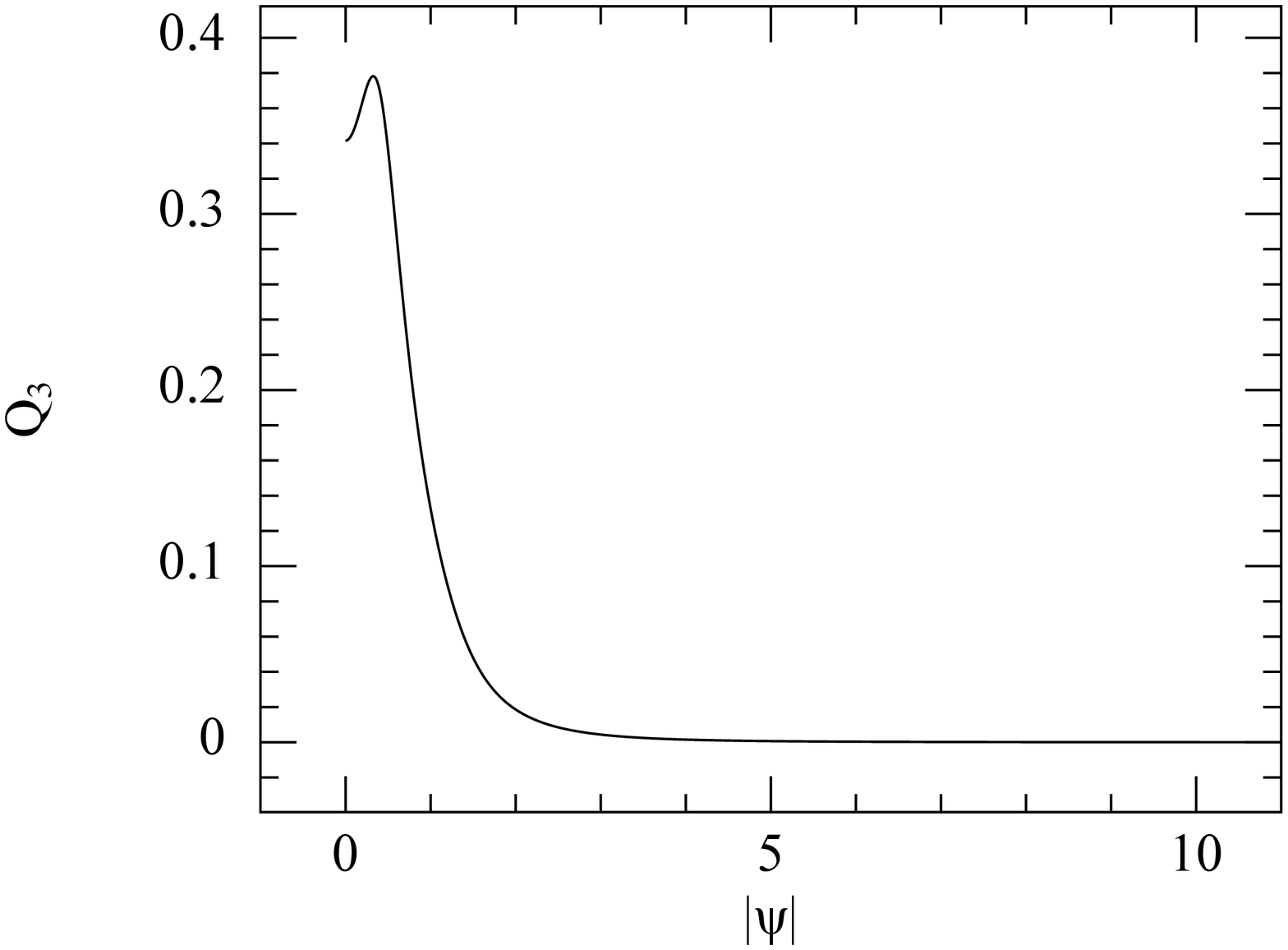}}
    \caption{The effective viscosity coefficients $Q_1$, $Q_2$ and $Q_3$
      plotted against warp amplitude $\left|\psi\right|$, assuming that
      $\alpha=0.2$, $\alpha_b=0$ and $\Gamma=1$. The coefficients change
      significantly for $0<\left|\psi\right|\lesssim 5$, after which they are
      approximately constant at a small but non-zero value}
    \label{Qs}
\end{figure*}

\subsection{Evolution equation}

The problem of disc breaking is complex, and to make progress we adopt the
simplest approach. We use a 1D Eulerian ring code of the type described in
\cite{Pringle1992} and used in \cite{LP2006}. Our code has been adapted to
include the third effective viscosity and the nonlinear fluid dynamical
constraints found by \citet{Ogilvie1999}. Here `1D' means that the properties
assigned to each ring depend only on its radius. Section~2.1 of \citet{LP2006}
gives a detailed discussion of the validity of this approach. The main
restriction is that the evolution of a disc warp must be diffusive rather than
wave--like.  \citet{PP1983} show that this holds if $\alpha > H/R$, which is
usually satisfied for accretion discs around compact objects.

Our 1D ring code describes the evolution of warped accretion discs by
evolving the angular momentum density vector of each ring
$\mathbi{L}(R,t)$ through the equation
\begin{eqnarray}
\label{dLdttotal}
 \frac{\partial \mathbi{L}}{\partial t} & = & \frac{1}{R}
 \frac{\partial }{\partial R} \left\{ \frac{\left(\partial / \partial
   R \right) \left[\nu_{1}\Sigma R^{3}\left(-\Omega^{'} \right)
     \right] }{\Sigma \left( \partial / \partial R \right) \left(R^{2}
   \Omega \right)} \mathbi{L}\right\} \\ \nonumber & &
 +~~{}\frac{1}{R}\frac{\partial}{\partial R}\left[\frac{1}{2} \nu_{2}R
   \left| \mathbi{L} \right|\frac{\partial \boldsymbol{\ell}}{\partial
     R} \right] \\ \nonumber & &
 +~~{}\frac{1}{R}\frac{\partial}{\partial R}
 \left\{\left[\frac{\frac{1}{2}\nu_{2}R^{3}\Omega \left|\partial
     \boldsymbol{\ell} / \partial R \right| ^{2}}{\left( \partial /
     \partial R \right) \left( R^{2} \Omega \right)} +
   \nu_{1}\left(\frac{R \Omega^{'}}{\Omega} \right) \right]
 \mathbi{L}\right\} \\ \nonumber & &
 +~~{}\frac{1}{R}\frac{\partial}{\partial R} \left(\nu_{3} R
 \left|\mathbi{L} \right| \boldsymbol{\ell} \times \frac{\partial
   \boldsymbol{\ell}}{\partial R} \right) \\ \nonumber & &
 +~~{}\boldsymbol{\Omega}_{\mathrm{p}} \times \mathbi{L}
\end{eqnarray}
where $\nu_1$, $\nu_2$ and $\nu_3$ are the effective viscosities,
$\Omega\left(R\right)$ is the local azimuthal angular velocity,
$\Sigma\left(R\right)$ is the disc surface density,
$\boldsymbol{\ell}\left(R\right)$ is the unit angular momentum vector, and
$\boldsymbol{\Omega}_{\rm p}\left(R\right)$ is the precession frequency
induced in the disc by the LT effect (defined below). We note that $\mathbi{L}
= \Sigma R^2 \Omega \boldsymbol{\ell}$.

Equation (\ref{dLdttotal}) shows that there are five independent torques
acting on the rings of gas. The first four terms on the rhs represent the
internal disc torques responsible for communicating angular momentum. The last
term represents the LT torque. Here we briefly discuss these terms.

The first term on the rhs of (\ref{dLdttotal}) describes the usual
viscous diffusion of mass. This is governed by the azimuthal shear
viscosity $\nu_{1}$. The second term is also diffusive. This term,
governed by the vertical viscosity $\nu_{2}$, is responsible for
diffusing the disc tilt. The third term is an advective
torque. Depending on its sign this advects angular momentum (and
hence mass) inwards or outwards through the disc.

The three terms discussed above are exactly those derived
straightforwardly from the conservation of mass and angular momentum
\citep{Pringle1992}. The fourth term is a precessional torque found by
the analysis in \citet{Ogilvie1999}.  This torque causes rings to
precess about $\boldsymbol{\ell} \times \frac{\partial
  \boldsymbol{\ell}}{\partial R}$ and thus only acts when the disc is
tilted. This leads to a dispersive wave--like propagation of the warp
\citep{Ogilvie1999}. The direction of the precession is dependent on
$\alpha$ and $\left|\psi\right|$ \citep[cf. Figure 5
  in][]{Ogilvie1999}. This term has exactly the same form as the
second term in (\ref{dLdttotal}) with the substitutions $0.5\nu_{2}
\to \nu_{3}$ and $\partial \boldsymbol{\ell}/ \partial R \to
\boldsymbol{\ell} \times \partial \boldsymbol{\ell}/\partial R$. This
transformation shows that
%this torque diffuses the disc twist similarly to the way the second term
%diffuses the disc tilt. It is also clear that 
we can integrate this term using the technique described in
\citet{Pringle1992} for integrating the second term.

The fifth term is the LT torque arising from the gravitomagnetic
interaction of the disc with the spinning hole. This induces
precessions in the disc orbits, and hence causes the angular momentum
vector for each ring to precess around the spin vector for the black
hole with a frequency
\begin{equation}
  \boldsymbol{\Omega}_{\rm{p}} = \frac{2G \mathbi{J}_{h}}{c^{2}R^{3}}
\label{omegap}
\end{equation} 
with $c$ the speed of light and $G$ the gravitational constant. The magnitude
of $\mathbi{J}_{\rm h}$ is given by $J_{\rm h} = acM(GM/c^{2})$ where $a$ is
the dimensionless spin parameter and $M$ is the mass of the hole
\citep{KP1985}. We include the back--reaction on the hole angular momentum
vector $\mathbi{J}_{\rm h}$ by summing the effect of the precessions over
all radii, giving the torque on the hole as:
\begin{equation}
  \frac{{\rm d} \mathbi{J}_{\rm h}}{{\rm d} t} = -2\pi \int
  \boldsymbol{\Omega}_{\rm p} \times \mathbi{L} R {\rm d}R.
\label{djstar}
\end{equation} 
Since $\boldsymbol{\Omega}_{\rm p} \propto \mathbi{J}_{\rm h}$ this equation
makes it clear that the magnitude of $\mathbi{J}_{\rm h}$ is conserved and
hence $\mathbi{J}_{\rm h}$ can only precess on a sphere, simply because the LT
torque has no component in the direction of $\mathbi{J}_{\rm h}$ for any disc
structure. Equivalently the LT torque induces only precessions in the disc, so
that the reaction back on the hole is simply a sum of precessions, which is
itself a precession.

We integrate both (\ref{dLdttotal}) and (\ref{djstar}) using a simple
forward Euler method \citep{Pringle1992} on a logarithmically spaced
grid. However we adopt a different implementation of the boundary
conditions. Rather than implementing a sink over a few grid cells we
simply remove all of the angular momentum that reaches the inner-- and
outer--most cells at the end of each integration step. This allows
angular momentum (and hence mass) to flow freely across the boundary,
giving the usual $\Sigma = 0$, torque--free, accreting boundary
condition. This is the form adopted in \citet{LP2010}. We note that
this boundary condition does not conserve angular momentum
exactly. However this small effect does not affect the inclination of
any of the rings and hence does not alter our results while the warp
is propagating through the computational domain. Once the warp reaches
the outer boundary it flows freely off the computational domain and
the disc is flattened (i.e. we do not impose a boundary condition at
the outer edge of the disc to maintain the disc warp).

\begin{figure*}
  \centering 
    \subfigure{\includegraphics[angle=0,width=0.33\textwidth]
       {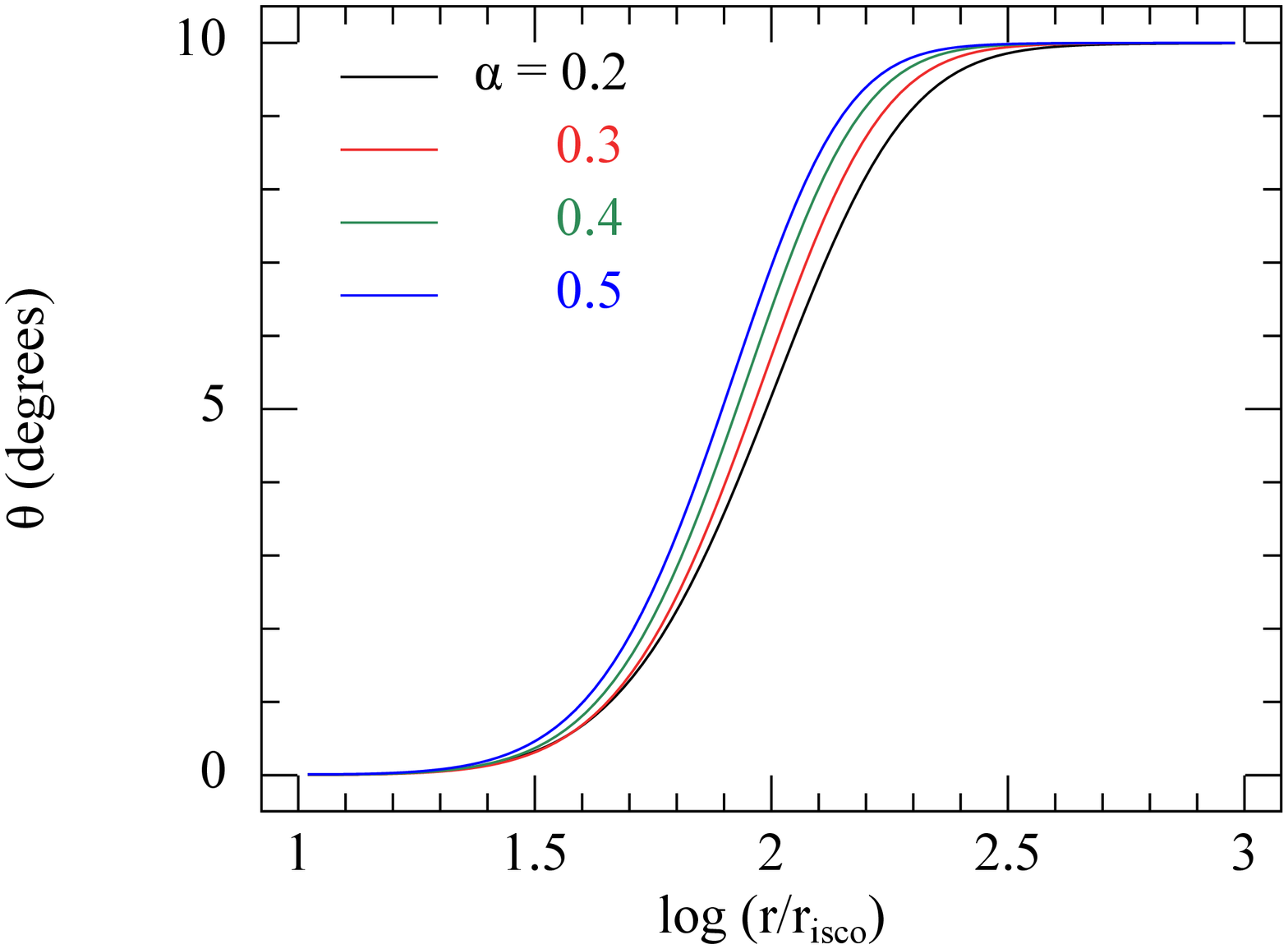}}
    \subfigure{\includegraphics[angle=0,width=0.33\textwidth]
       {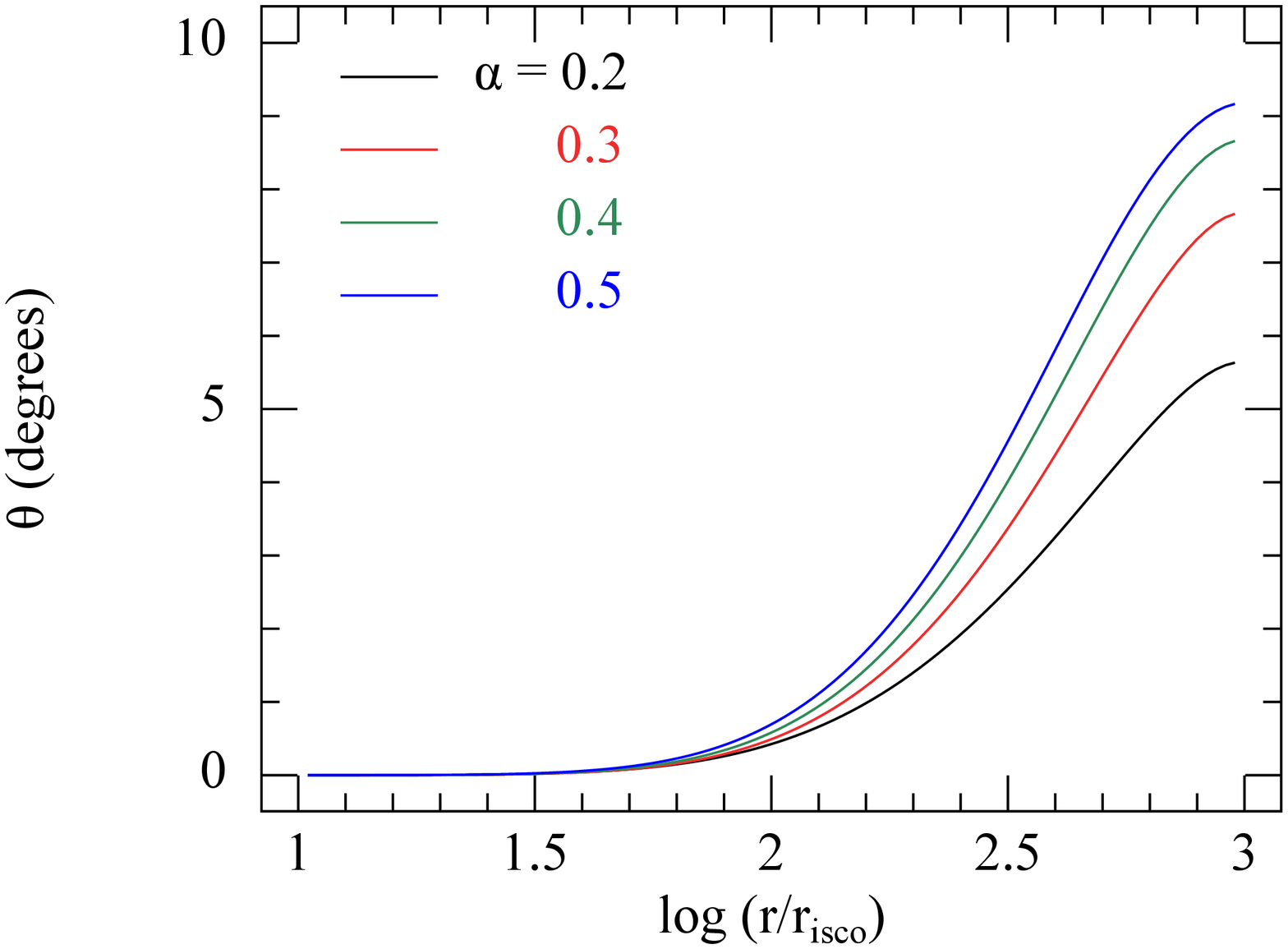}}
    \subfigure{\includegraphics[angle=0,width=0.33\textwidth]
       {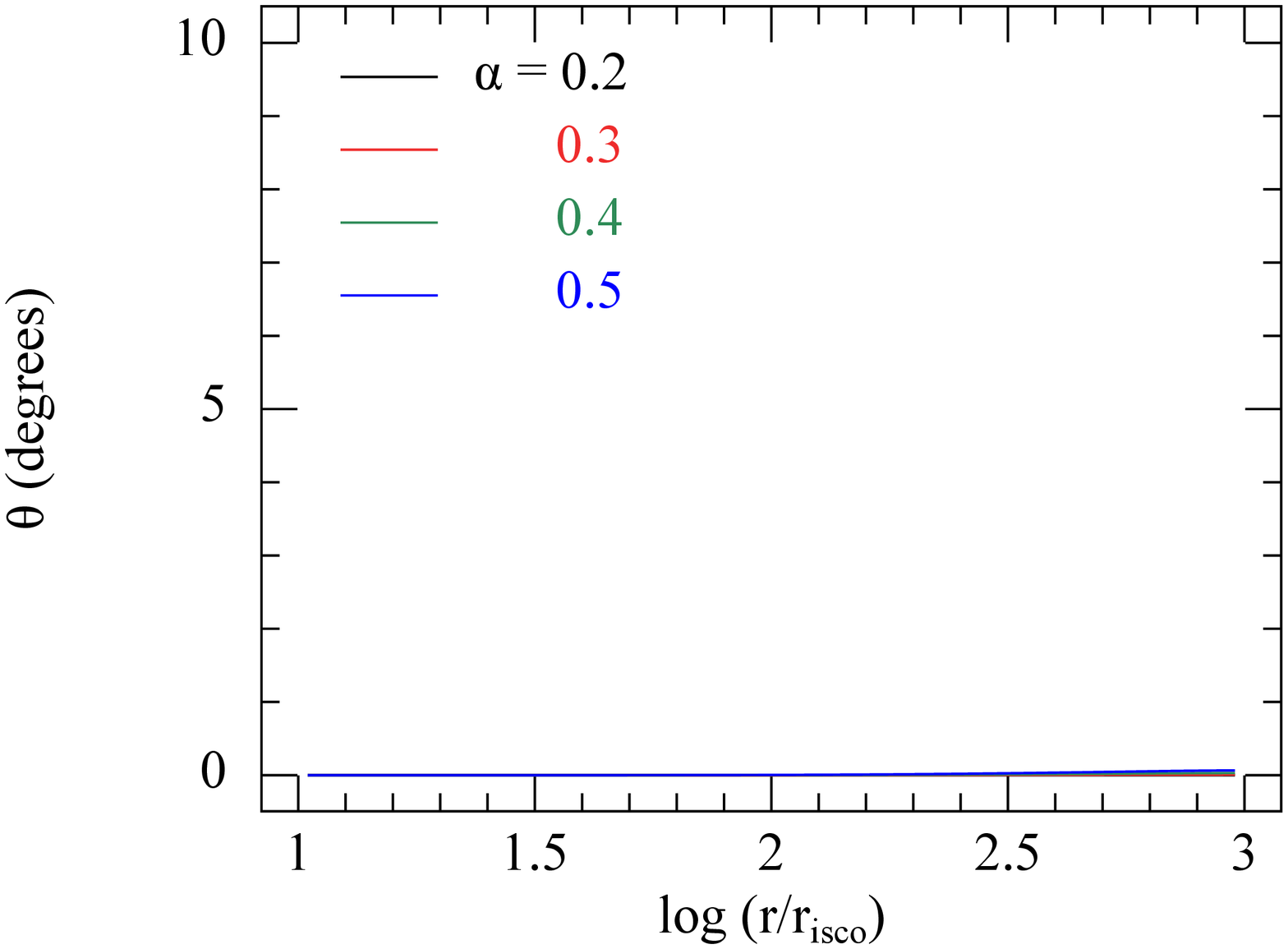}}
    \caption{Disc structures for the constant effective viscosity simulations,
      given by plotting the tilt angle ($\theta$) between the disc and the
      hole against the log of the radius. The initial misalignment of the disc
      is $\theta_0 = 10^{\circ}$. On each plot the legend gives the value of
      $\alpha$ used in the simulation. From left to right the panels
      correspond to $t = 0.01$, $0.1$ and $1$ viscous times after the start of
      the calculation. Note that in the latter case the warp has moved to the
      edge of our grid and the disc aligns.}
    \label{const10}
\end{figure*}
\begin{figure*}
  \centering
    \subfigure{\includegraphics[angle=0,width=0.33\textwidth]
       {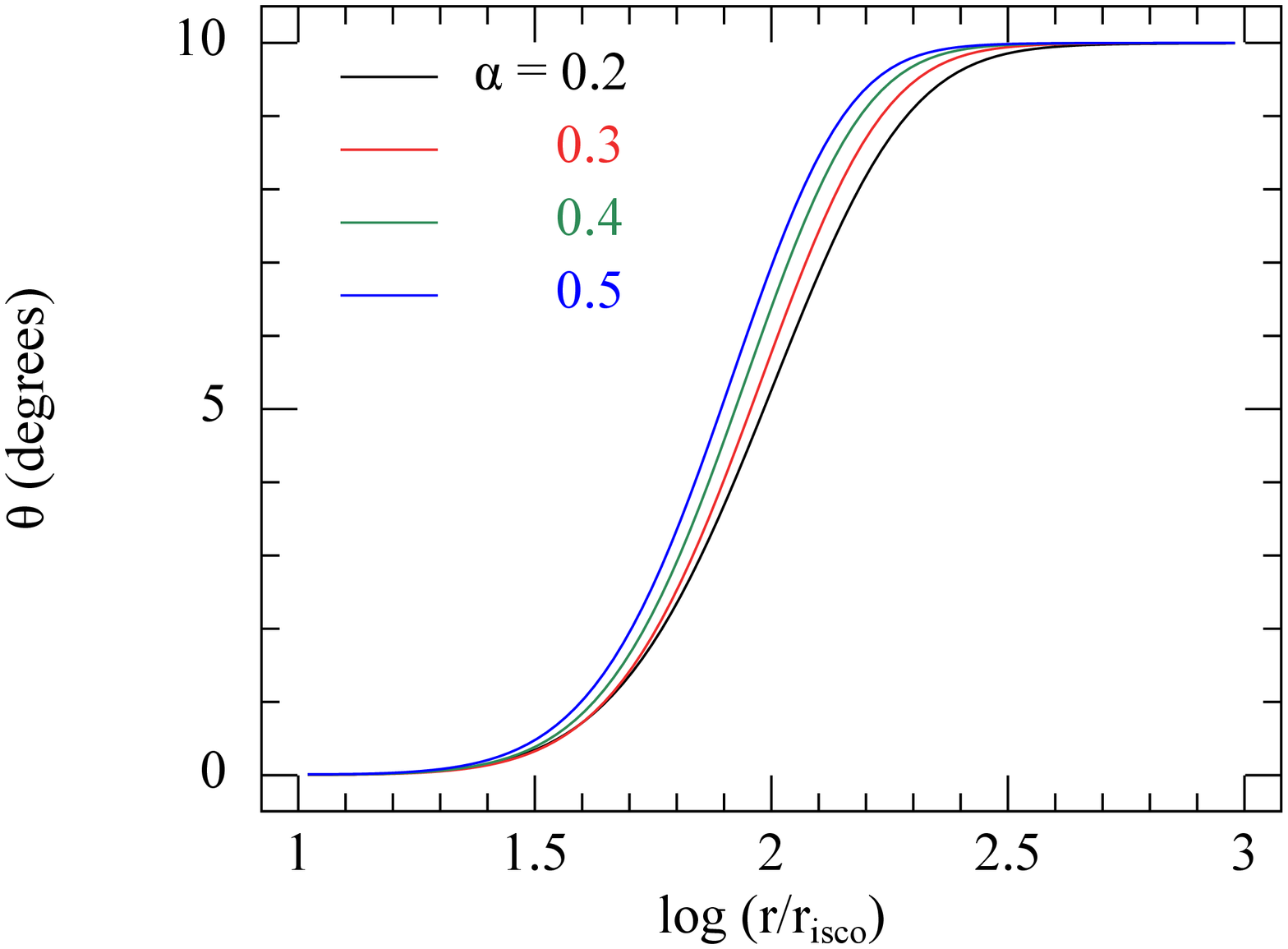}}
    \subfigure{\includegraphics[angle=0,width=0.33\textwidth]
       {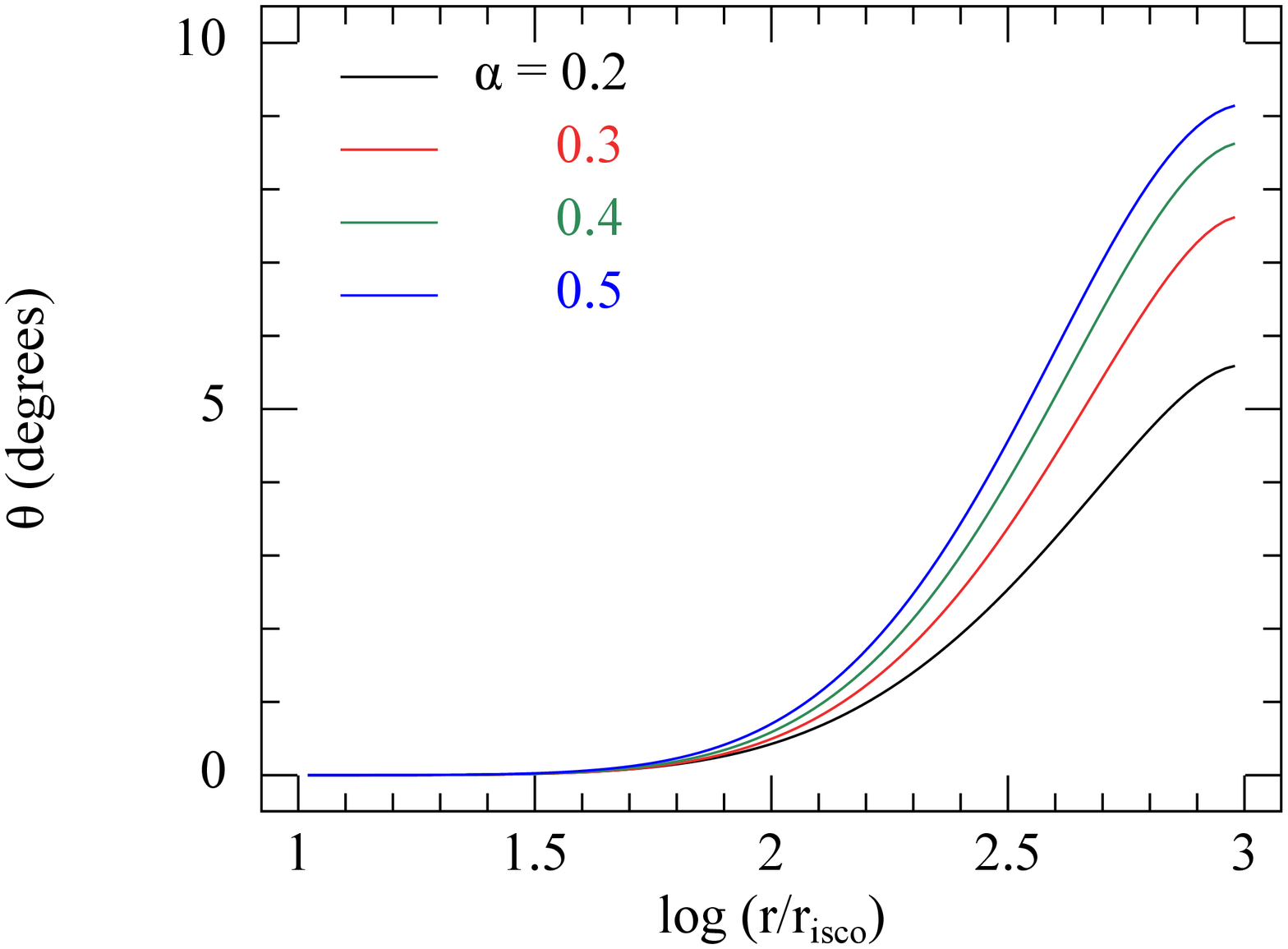}}
     \subfigure{\includegraphics[angle=0,width=0.33\textwidth]
       {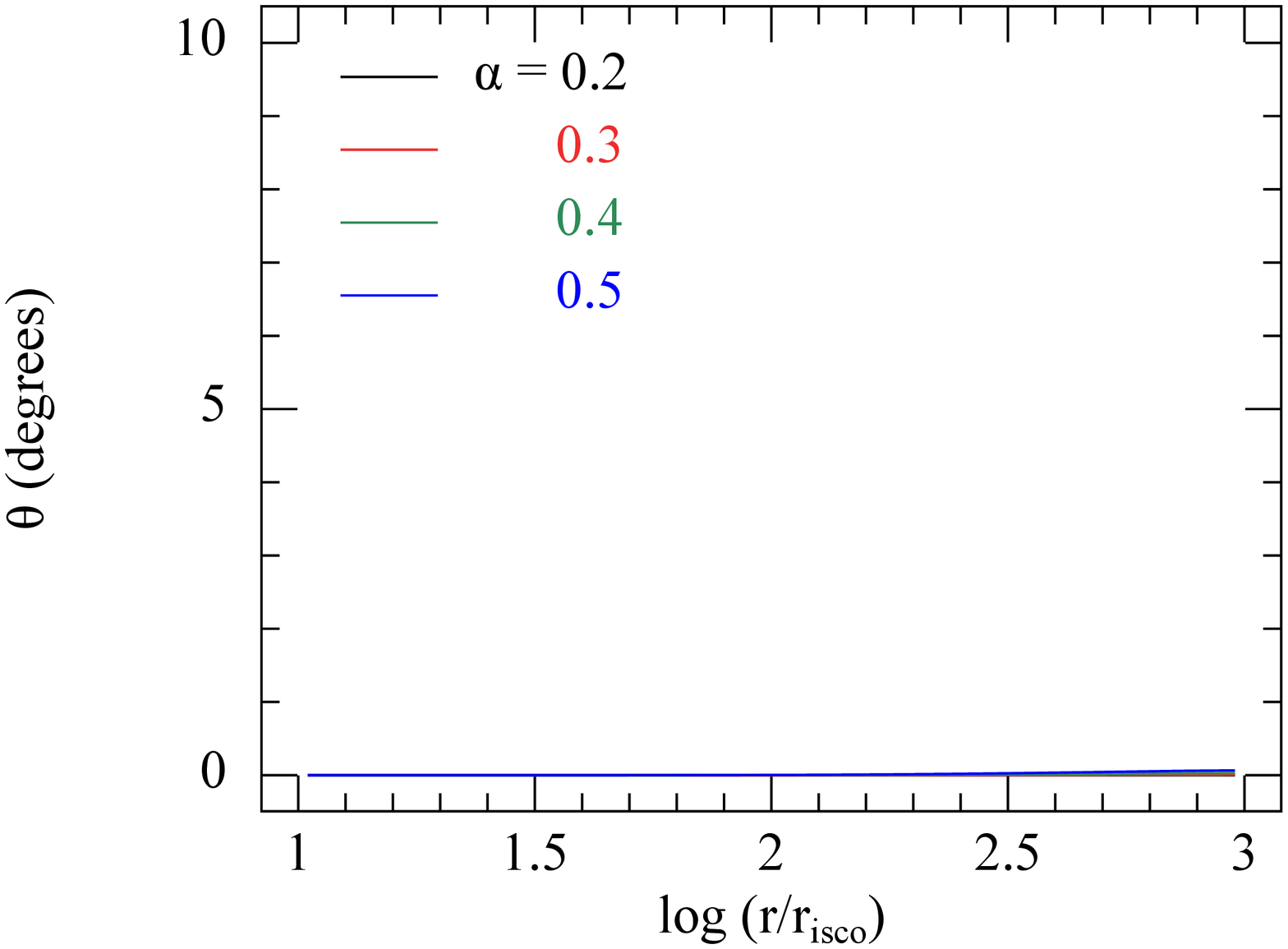}}
   \caption{Disc structures for the full effective viscosity simulations,
     given by plotting the tilt angle between the disc and the hole against
     the log of the radius. The initial misalignment of the disc is $\theta_0
     = 10^{\circ}$. On each plot the legend gives the value of $\alpha$ used
     in the simulation. From left to right the panels correspond to $t =
     0.01$, $0.1$ and $1$ viscous times after the start of
      the calculation. When compared to \fref{const10} this
     figure shows the negligible effect of the full effective viscosities when
     the misalignment between the disc and hole is small.}
    \label{og10}
\end{figure*}

\subsection{Generating the constrained effective viscosities}
\label{Ogvisc}
We use the effective viscosities derived by \citet{Ogilvie1999} for a locally
isothermal disc. These are constrained to be consistent with the internal
fluid dynamics of the disc, assuming an isotropic physical viscosity. The
physics of these constraints is simply local conservation of mass and angular
momentum in the disc.

The effective viscosities which describe the transport of angular
momentum in a warped disc take the usual form of a Shakura \& Sunyaev
disc viscosity. The constraints determine the nonlinear coefficients
($Q_1, Q_2, Q_3$) of the effective viscosities (cf equations
\ref{nu1final}, \ref{nu2final} and \ref{nu3final}). These coefficients
all decrease in magnitude in the presence of a strong warp, where the
disc rings are highly inclined to each other. Communication of angular
momentum is reduced because the rings of gas are no longer in perfect
contact. We expect angular momentum to be communicated through
perturbations of the particles from circular orbits. The rate of
communication, and hence the magnitude of the viscosities, therefore
depend on how easily these particles can interact. This is harder for
particles on inclined orbits, and once these reach a critical
inclination this can only happen where the rings cross.  We note that
for inviscid discs, which we do not consider here, this interpretation
is no longer valid, see \citet{Ogilvie1999} for further details.

We now detail the calculation of the effective viscosities from the
disc properties. The notation in \cite{Ogilvie1999} differs from that
in \citet{Pringle1992}. Our \eref{dLdttotal}, and equation~122 from
\citet{Ogilvie1999} relate the effective viscosities ($\nu_1$, $\nu_2$
and $\nu_3$) to the nonlinear viscosity coefficients ($Q_1$, $Q_2$
and $Q_3$) by
\begin{equation}
\nu_{1} = \frac{Q_{1} \mathscr{I} \Omega^{2}}{\Sigma R
  \frac{\mathrm{d}\Omega}{\mathrm{d}R}},
\label{nu1Q1}
\end{equation}
\begin{equation}
\nu_{2} = \frac{2 Q_{2}\mathscr{I} \Omega}{\Sigma}
\label{nu2Q2}
\end{equation}
and
\begin{equation}
\nu_3 = \frac{Q_{3}\mathscr{I} \Omega}{\Sigma},
\label{nu3Q3}
\end{equation}
where $\mathscr{I}$ is given by
\begin{equation}
\mathscr{I} = \frac{1}{2 \pi} \int^{2\pi}_{0}{\tilde{\mathscr{I}}}
~\mathrm{d}\phi = \frac{1}{2 \pi} \int^{2\pi}_{0} \int^{\infty}_{-\infty}\rho
z^{2}~\mathrm{d}z ~\mathrm{d}\phi.
\label{I}
\end{equation}
Assuming $\Sigma$ constant in azimuth, and a locally isothermal disc
with sound speed $c_s$, we can write this as $\mathscr{I} \propto \Sigma
c_s^2/\Omega^2$. For a point--mass potential
with $\Omega = (GM/R^{3})^{1/2}$ we get after some algebra
\begin{equation}
\nu_{1} = -\frac{2}{3}Q_{1}\left[\left(H/R\right)^{2}R^{2}\Omega \right],
\label{nu1final}
\end{equation}
\begin{equation}
\nu_{2} = 2Q_{2}\left[\left(H/R\right)^{2}R^{2}\Omega \right]
\label{nu2final}
\end{equation}
and
\begin{equation}
  \nu_{3} = Q_{3} \left[ \left(H/R \right)^{2} R^{2}\Omega \right],
\label{nu3final}
\end{equation}
where $H = c_{\rm s}/\Omega$ is the disc thickness.

For simplicity we assume that
\begin{equation}
\left(H/R\right)^{2}R^{2}\Omega = \rm{const}.
\label{const}
\end{equation}
This removes the radial dependence of the viscosities, making them functions
only of $\left|\psi\right|$ and $\alpha$ and so allowing a direct comparison
with the constant viscosity simulations of \citet{LP2006}. The assumption
(\ref{const}) requires that the sound speed $c_{\rm s} \propto R^{-3/4}$. This
compares with the usual steady--state disc, where $c_{\rm s} \propto R^{-3/8}$
(e.g. \citealt{Franketal2002}).

We note that for a disc without a warp, $\nu_{1}$ reduces to exactly
the usual $\alpha$ disc viscosity \citep{SS1973}, since
\begin{equation}
  \nu_{1} = \alpha c_{\rm s} H = \alpha \left(H/R\right)^{2} R^{2}\Omega 
\label{nu1ss}
\end{equation} 
and 
\begin{equation}
  Q_{1}(\left|\psi\right|=0) = -3\alpha/2.
\label{Q10}
\end{equation} 

The effective viscosities are now fixed by the nonlinear coefficients
$Q_1\left(\alpha,\left|\psi\right|\right)$,
$Q_2\left(\alpha,\left|\psi\right|\right)$ and
$Q_3\left(\alpha,\left|\psi\right|\right)$.  We calculate these
coefficients in the same way as described in \citet{Ogilvie1999} with
$\Gamma=1$ and $\alpha_b=0$ using the code kindly provided by Gordon
Ogilvie. Figure~\ref{Qs} show the coefficients plotted against the
warp amplitude for $\alpha=0.2$.
 
%%%%%%%%%%%%%%%%%%%%%%%%%%%%%%%%%%%%%%%%%%%%%%%%%%%%%%%%%%%%%%%%%%%%%%%%%%%%%%
%% Numerical Simulations                                                    %%
%%%%%%%%%%%%%%%%%%%%%%%%%%%%%%%%%%%%%%%%%%%%%%%%%%%%%%%%%%%%%%%%%%%%%%%%%%%%%%
\section{Simulations}
\label{simulations}

We use the method described above to simulate the warping of a
misaligned accretion disc under the LT effect. We set up our initially
planar but misaligned disc on a logarithmically spaced grid from
$R_{\rm in}=10R_{\rm isco}$ to $R_{\rm out}=1000R_{\rm isco}$ using
100 grid cells (where $R_{\rm isco}=6GM/c^2$ is the radius of the
innermost stable circular orbit around a Schwarzchild black hole
). The disc surface density is initially a Gaussian ring peaked at
$R_{0} = 500R_{\rm isco}$ with width $50R_{\rm isco}$. We initialise
each ring with a constant misalignment angle, $\theta_0$, such that
the initial disc is flat but tilted to the plane of the black hole.

\citet{LP2006} suggest that the only governing parameters for simulations of
this type are the ratio of disc to hole angular momenta ($J_{\rm d}/J_{\rm
  h}$), the ratio of the warp radius $R_{\rm w} = \Omega_{\rm p} R^3 / \nu_2$
to $R_0$, and the ratio of the effective viscosities $\nu_2/\nu_1$ (note that
they neglect $\nu_3$). However, as we use the constrained effective
viscosities, $\nu_2/\nu_1$ is now determined once we have chosen $\alpha$.

The parameters that we are free to choose in these simulations are $\alpha$,
$H/R$, $R_0$, $\theta_0$ and $J_{\rm d}/J_{\rm h}$. For black hole accretion
discs $H/R \sim 10^{-3}$ and so we assume this throughout. \citet{LP2006}
explored the effect of changing $R_{\rm w}/R_0$, so we fix $R_0$ as above
(note that this does not fix the ratio $R_{\rm w}/R_0$. However this ratio
varies little throughout our simulations.) To examine disc breaking we choose
$J_{\rm d}/J_{\rm h} \ll 1$ as the simplest case and then vary $\alpha$ and
$\theta_0$.

\begin{figure*}
  \centering 
    \subfigure{\includegraphics[angle=0,width=0.33\textwidth]
       {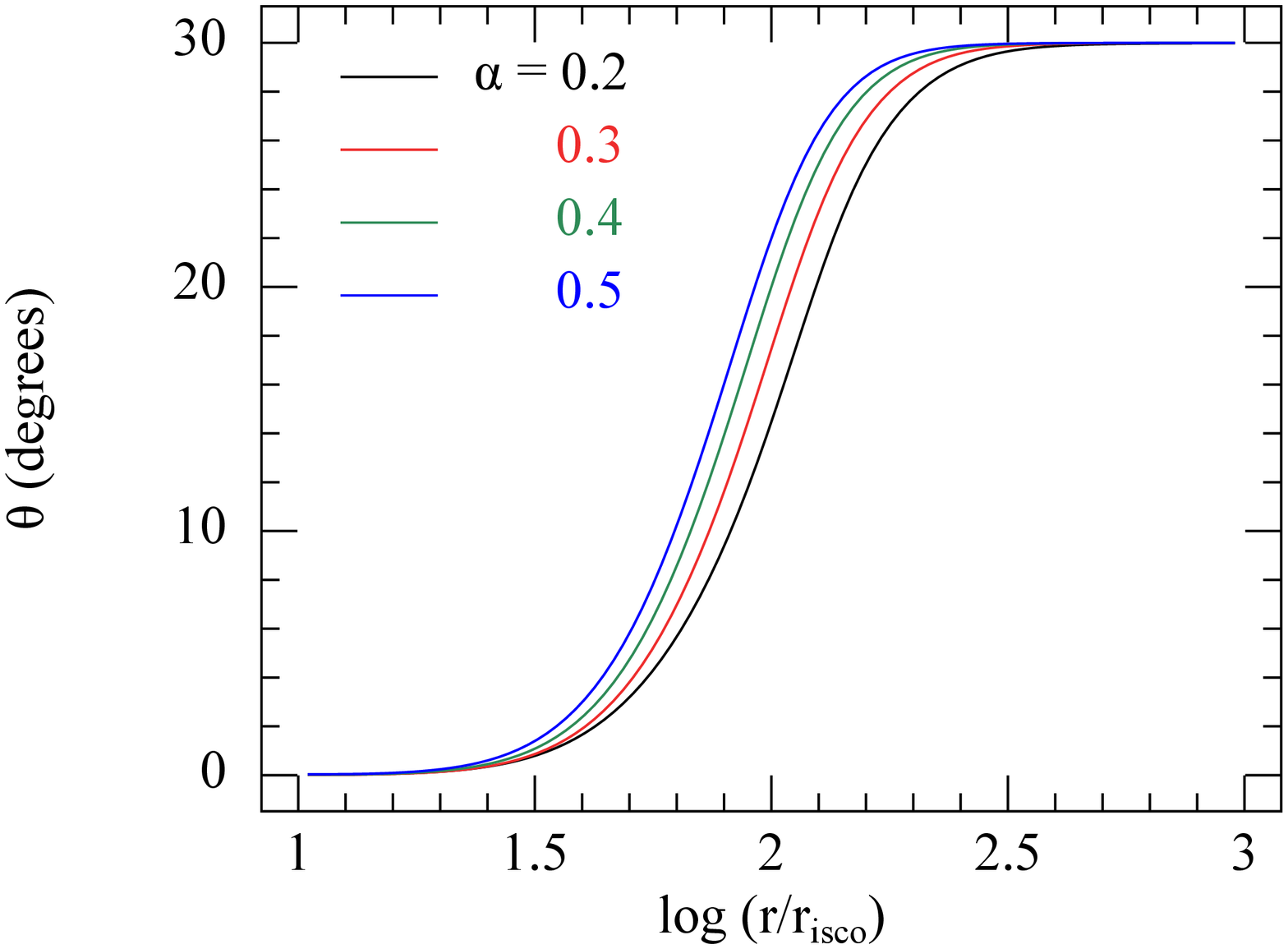}}
    \subfigure{\includegraphics[angle=0,width=0.33\textwidth]
       {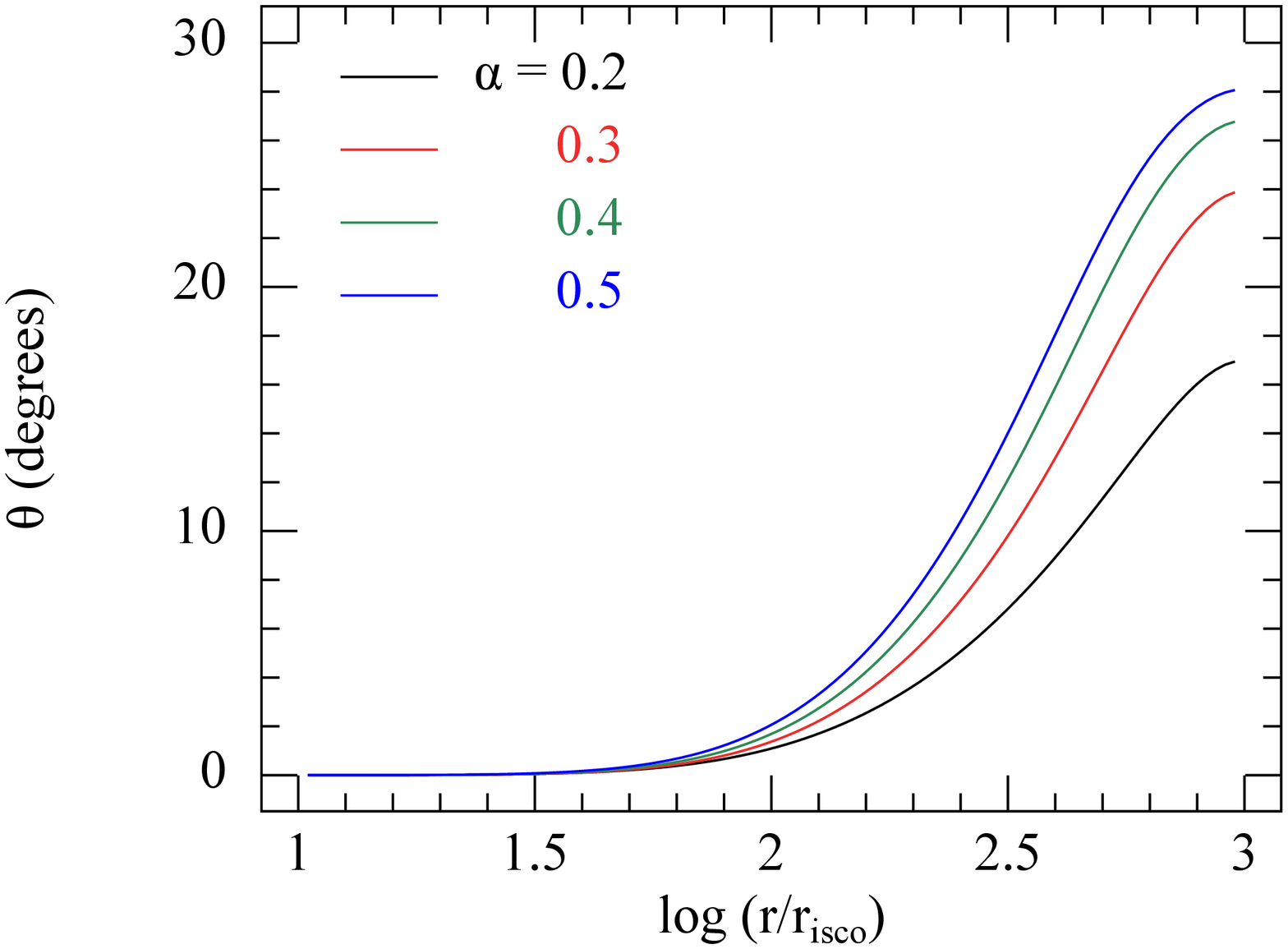}}
    \subfigure{\includegraphics[angle=0,width=0.33\textwidth]
       {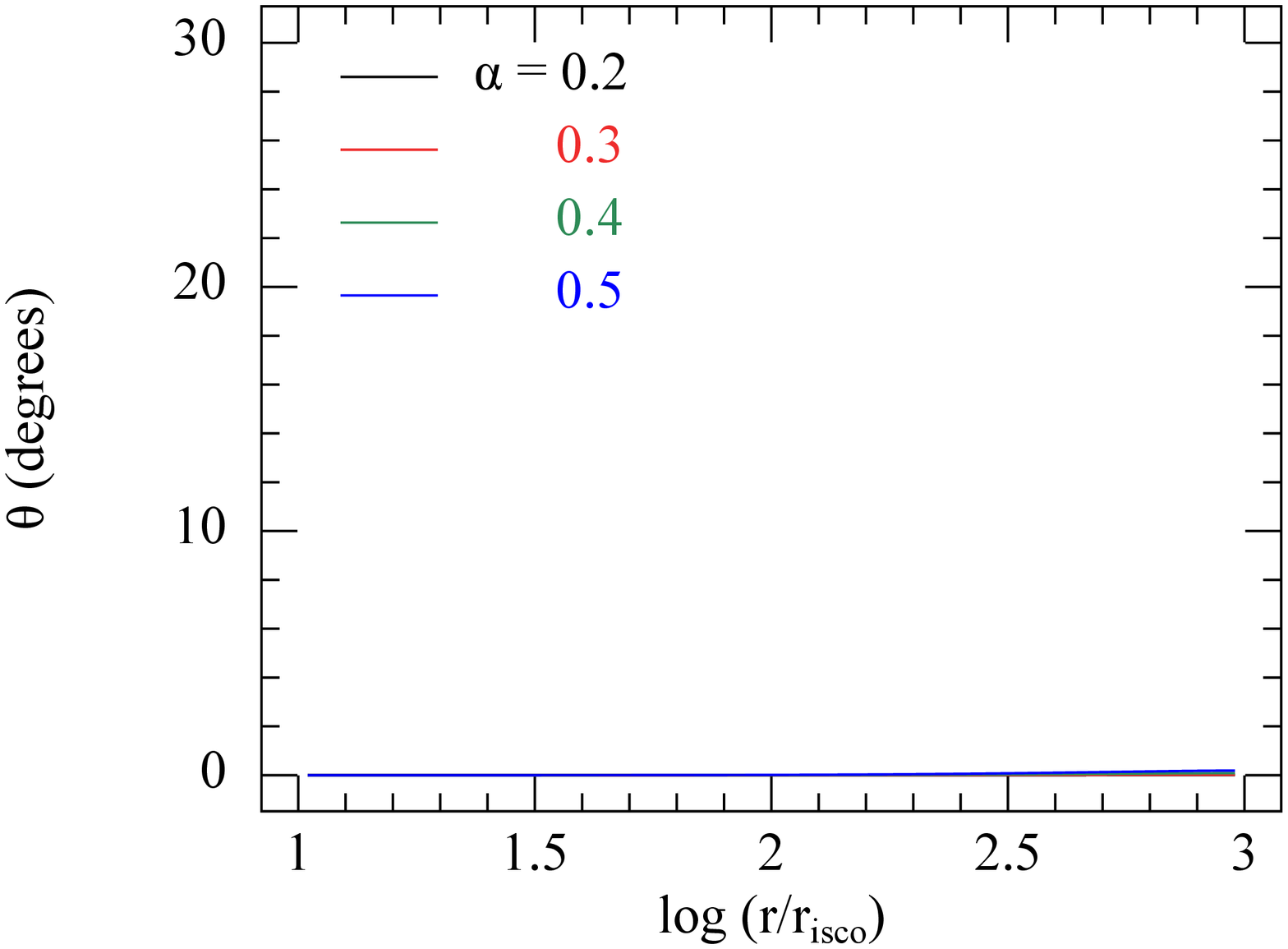}}
    \caption{Disc structures for the constant effective viscosity simulations
      as in Fig.~\ref{const10}, but this time for initial misalignment $\theta
      = 30^{\circ}$. From left to right the panels correspond to $t = 0.01$,
      $0.1$ and $1$ viscous times after the start of the calculation.}
    \label{const30}
\end{figure*}
\begin{figure*}
  \centering
    \subfigure{\includegraphics[angle=0,width=0.33\textwidth]
       {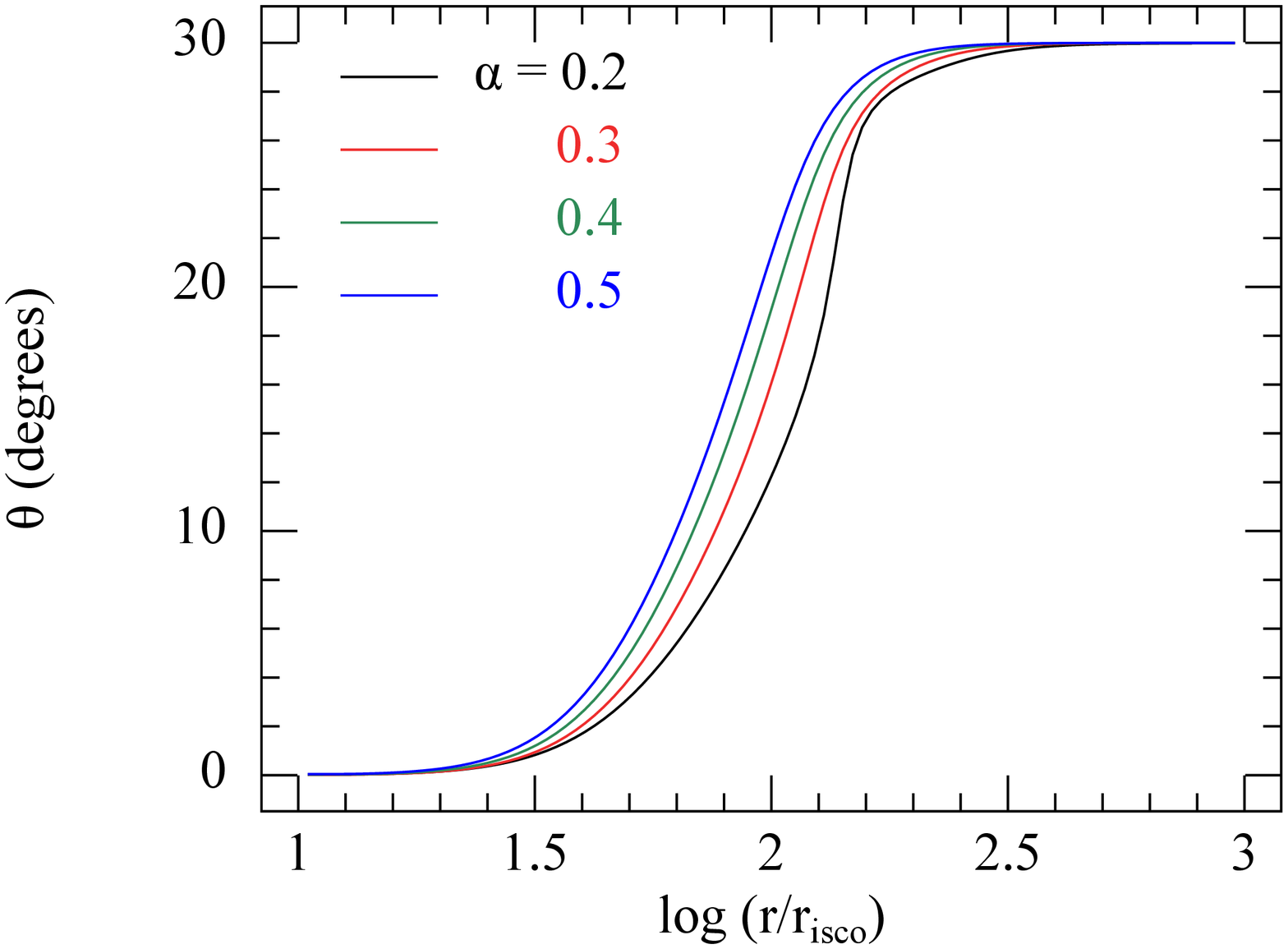}}
    \subfigure{\includegraphics[angle=0,width=0.33\textwidth]
       {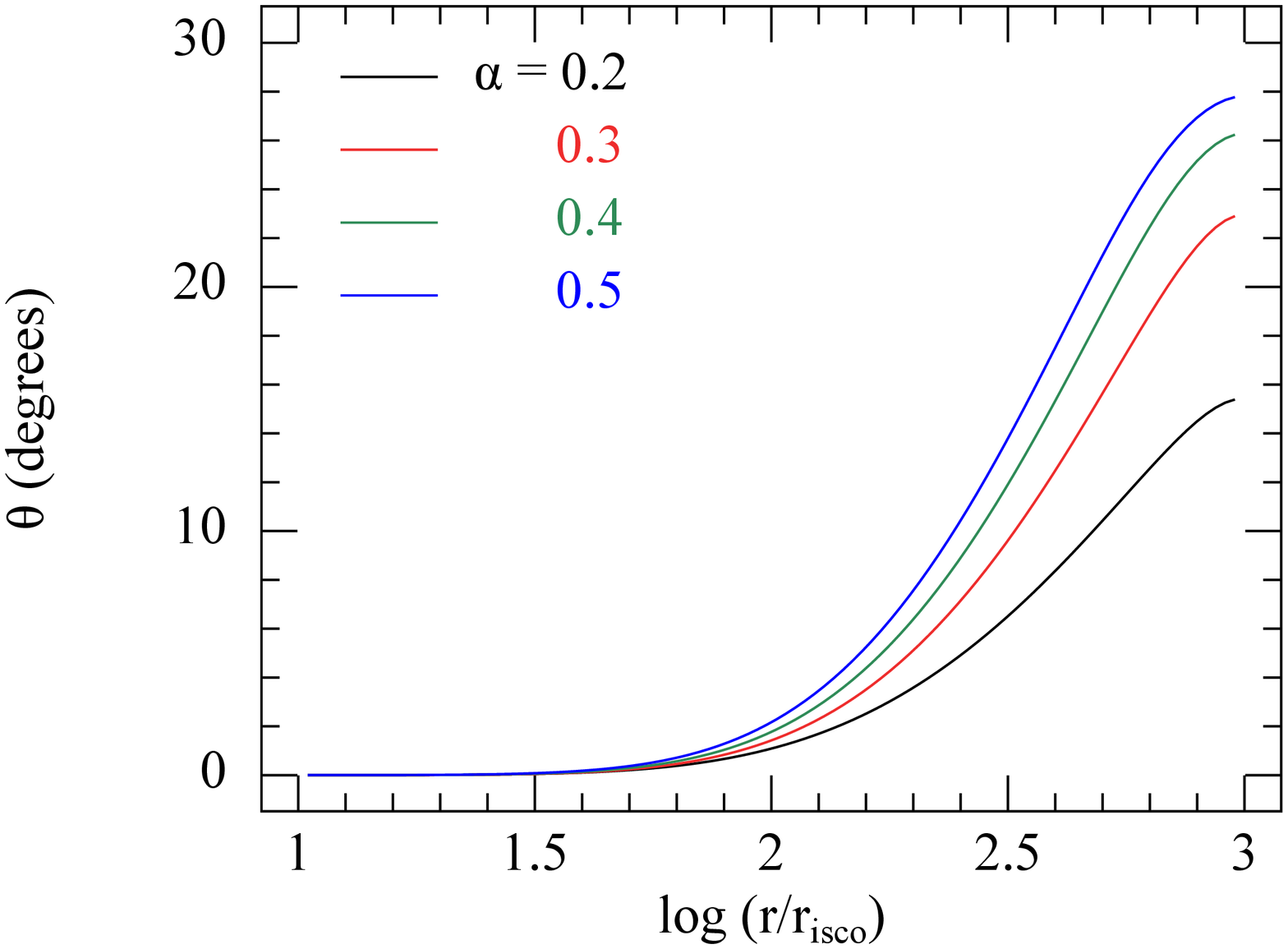}}
    \subfigure{\includegraphics[angle=0,width=0.33\textwidth]
       {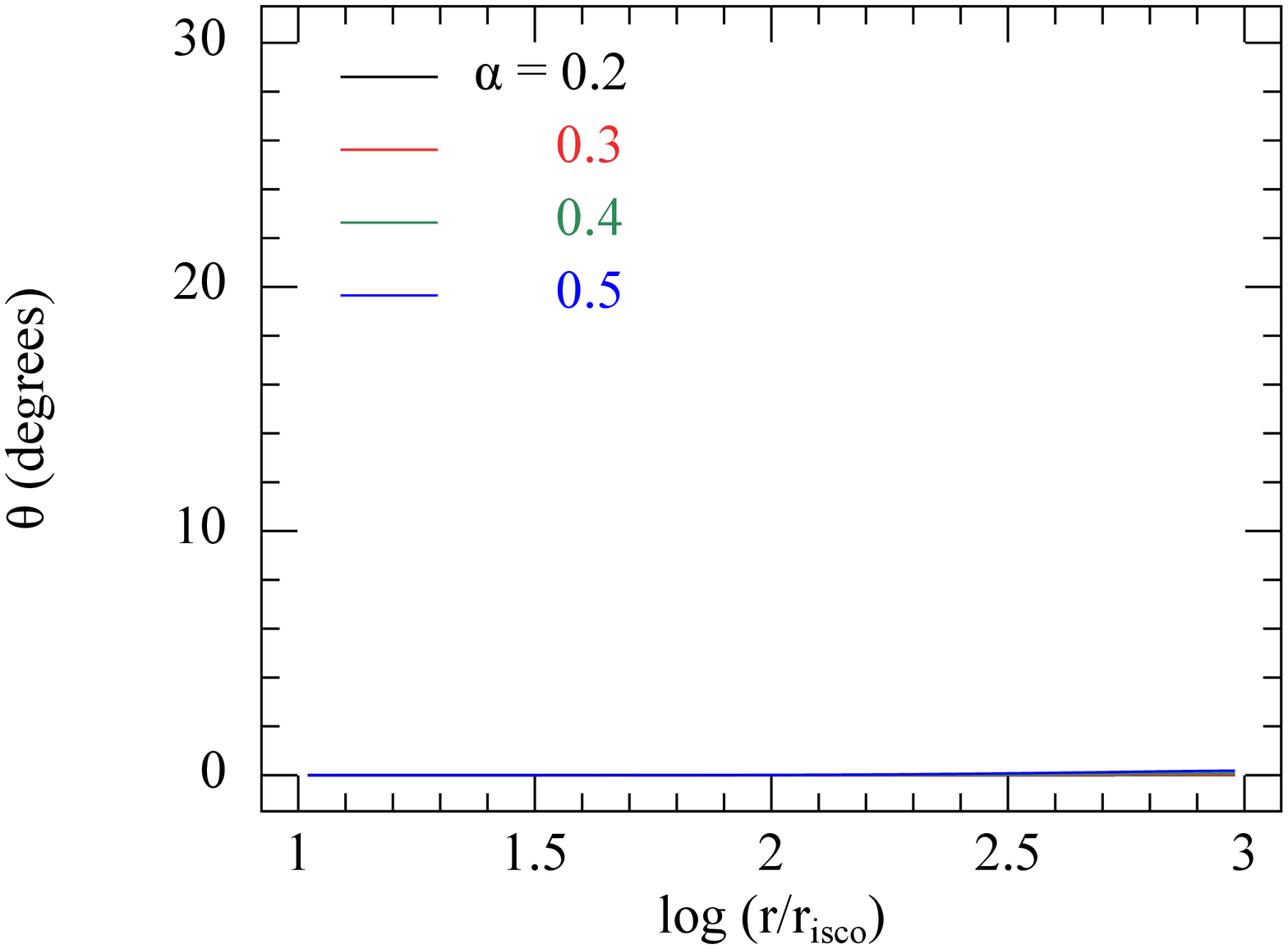}}
    \caption{Disc structures for the full effective viscosity simulations as
      in Fig.~\ref{og10}, but this time for initial misalignment $\theta =
      30^{\circ}$.  From left to right the panels correspond to $t = 0.01$,
      $0.1$ and $1$ viscous times after the start of the calculation. When
      compared to \fref{const30} this figure shows that even for a
      $30^{\circ}$ misalignment angle the effect of the full effective
      viscosities can be small, however there is a steepening of the disc
      profile at early times for the $\alpha=0.2$ simulations. This suggests
      that for the nonlinear effects to be important at $\theta_0 =
      30^{\circ}$ we need $\alpha \ll 0.2$.}
    \label{og30}
\end{figure*}

\subsection{Does the disc break?}
\label{breaking}
We perform several simulations for $0.2 \le \alpha \le 0.5$ to determine the
various nonlinear responses of the disc to warping. We do not simulate $\alpha
< 0.2$ as this is where the viscosity $\nu_1$ is predicted to become negative
in strong warps. We simulate a range of $\theta_0$ from $10^{\circ}$ to
$60^{\circ}$. This allows us to explore the two regimes: $\left|\psi\right|
\lesssim \alpha$, where nonlinear effects should be negligible, and
$\left|\psi\right| \gtrsim \alpha$, where nonlinear effects should be
important. Our setup corresponds to $J_{\rm d}/J_{\rm h} \approx 0.02$, with
$R_{\rm w}/R_{0}$ varying between $\sim 0.95$ and $\sim 1.20$ as $\alpha$
changes. Note that we evaluate (\ref{const}) at $R_0$ with $H/R = 10^{-3}$.

We perform these simulations for two cases. In the first we use constant
effective viscosities. This acts as a control so we can isolate the effect of
introducing nonlinear dynamics in the second set of simulations, where we use
the full effective viscosities. The constant effective viscosities are simply
calculated using the method described in Section~\ref{Ogvisc} with
$\left|\psi\right|=0$.

For the simulations involving constant effective viscosities we expect to
recover the Bardeen--Petterson effect as seen by several other authors
(e.g. \citealt{BP1975}; \citealt{Pringle1992}; \citealt{SF1996};
\citealt{LP2006} etc). However the simulations involving the full nonlinear
effective viscosities should modify this somewhat.

\subsubsection{$\theta = 10^{\circ}$}
\label{theta10}

In \fref{const10} we show the disc structures for simulations with
$\theta_0=10^{\circ}$ and the constant effective viscosities at $0.01$, $0.1$
and $1$ $t_{\rm visc}$. Here we define the viscous timescale as
$t_{\rm visc} = R_0^2/\nu_1$ with $\nu_1$ evaluated for $\left|\psi\right|=0$
and $\alpha$ as defined in the particular case. \fref{const10} shows the
behaviour for different values of the disc viscosity parameter $\alpha$. In
this case the behaviour is similar for all simulations, with lower $\alpha$
aligning in a shorter fraction of the viscous time (although in reality more
slowly because the viscous timescale is longer).

In \fref{og10} we show the same simulations as in \fref{const10}. However this
time we use the full effective viscosities. We expect the evolution here to be
similar, as the warp is never able to achieve a large amplitude, making the
viscosities similar in both cases. This is indeed the case, as there is no
visible difference between Figures~\ref{const10} \& \ref{og10}.

\subsubsection{$\theta = 30^{\circ}$}
\label{theta30}
In Figures~\ref{const30} \& \ref{og30} we show the disc structures for
simulations with $\theta_0=30^{\circ}$. Again there is very little difference
between the disc structures obtained using the constant effective viscosities
and those with the full effective viscosities. However the disc profile
steepens in the early stages of the $\alpha=0.2$ simulation. This suggests
that for $\theta=30^{\circ}$ we would have to go to $\alpha \ll 0.2$ to get
disc breaking.

\subsubsection{$\theta = 45^{\circ}$}
\label{theta45}
\begin{figure*}
  \centering 
    \subfigure{\includegraphics[angle=0,width=0.33\textwidth]
       {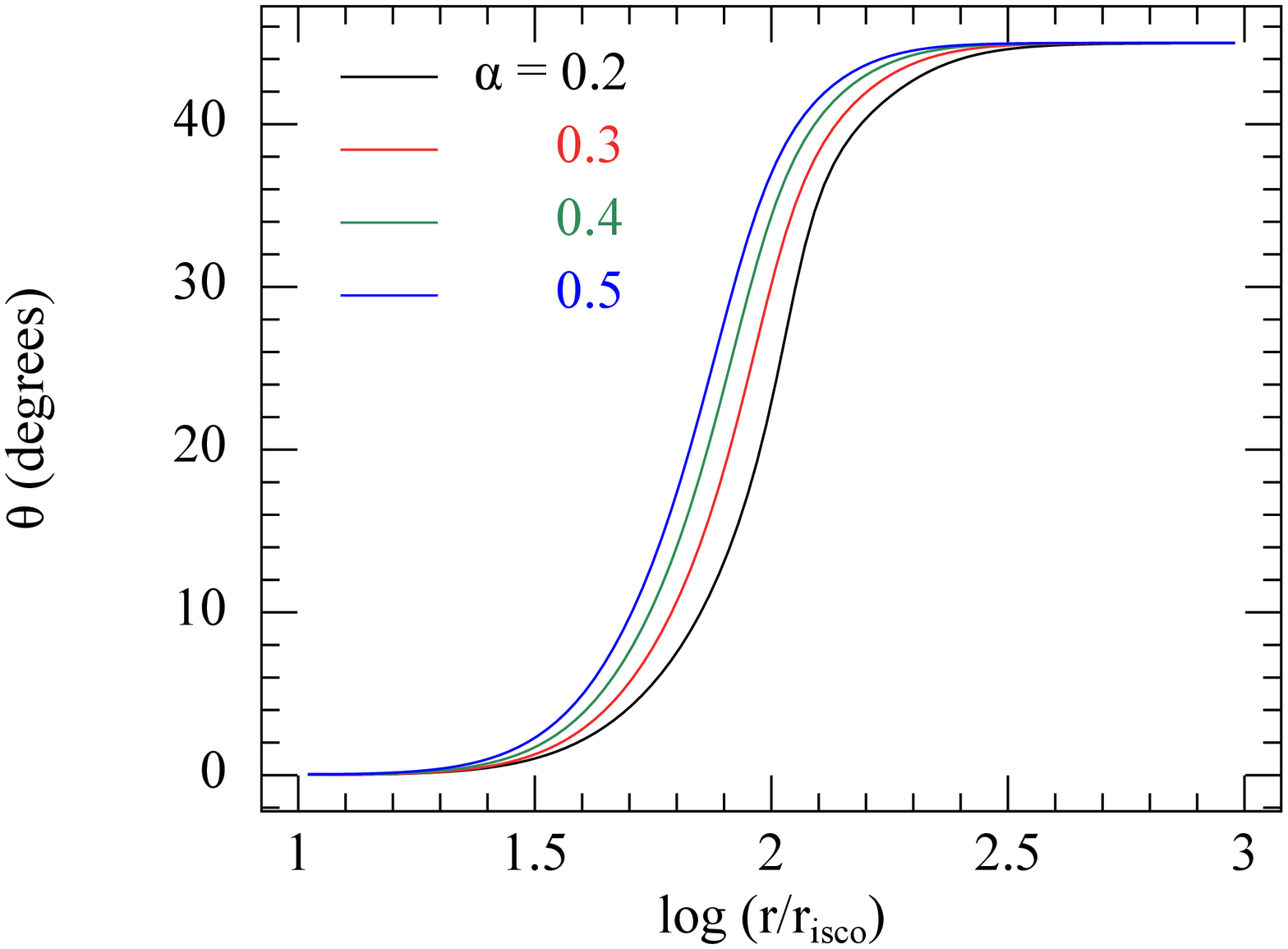}}
    \subfigure{\includegraphics[angle=0,width=0.33\textwidth]
       {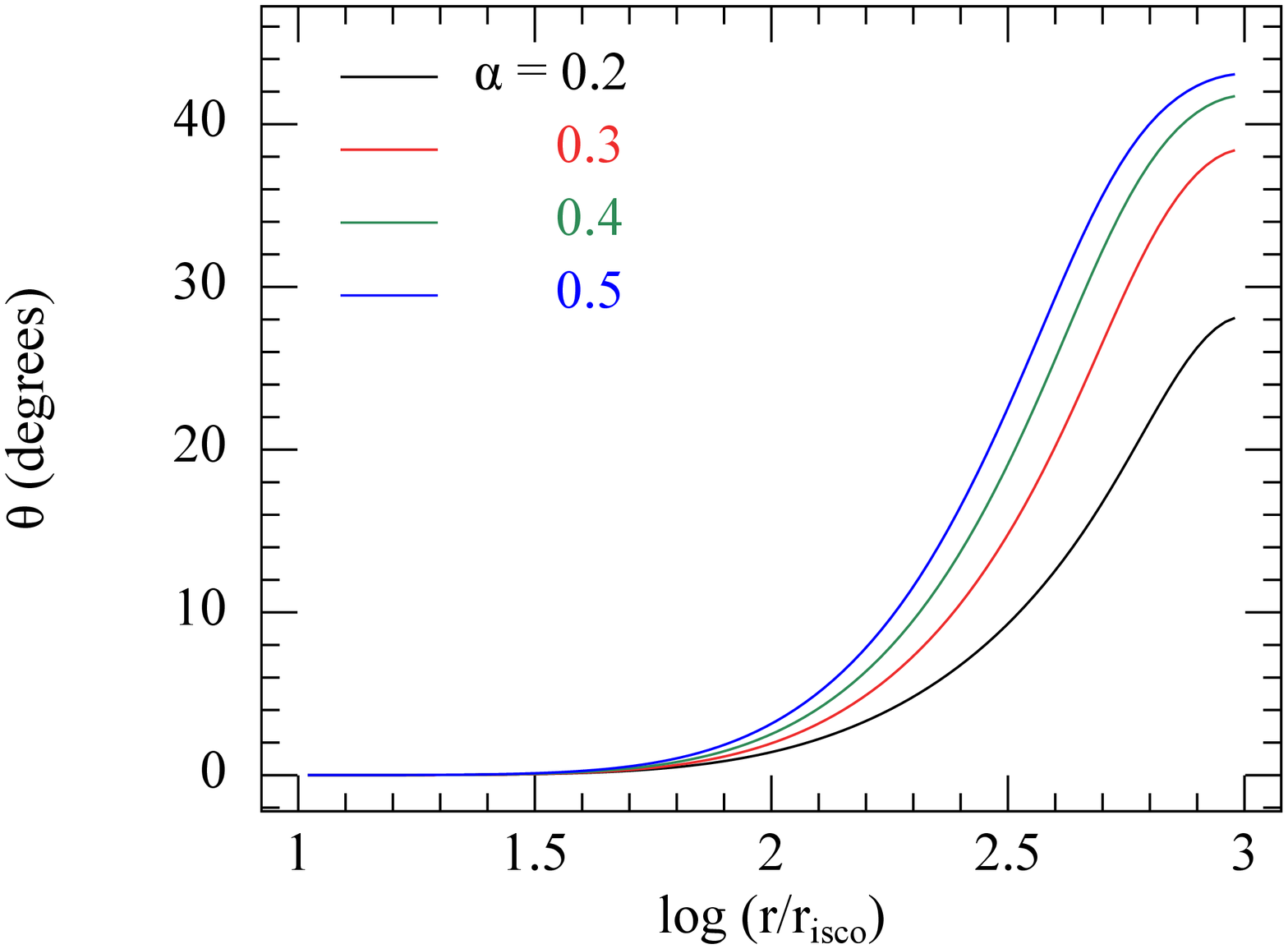}}
    \subfigure{\includegraphics[angle=0,width=0.33\textwidth]
       {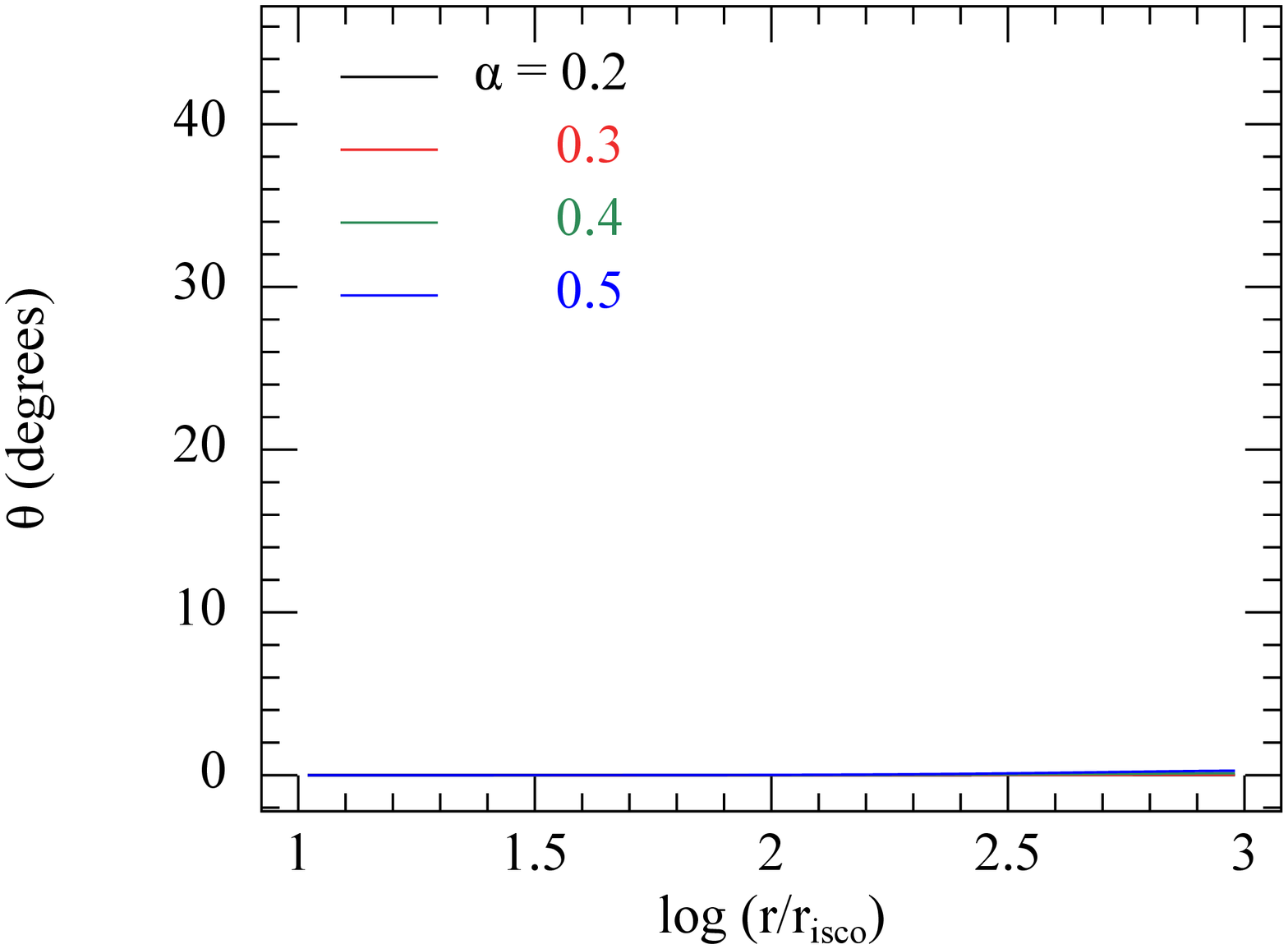}}
    \caption{Disc structures for the constant effective viscosity simulations
      as in Fig.~\ref{const10}, but this time for initial misalignment $\theta
      = 45^{\circ}$. From left to right the panels correspond to $t = 0.01$,
      $0.1$ and $1$ viscous times after the start of the calculation.}
    \label{const45}
\end{figure*}

\begin{figure*}
  \centering
    \subfigure{\includegraphics[angle=0,width=0.33\textwidth]
       {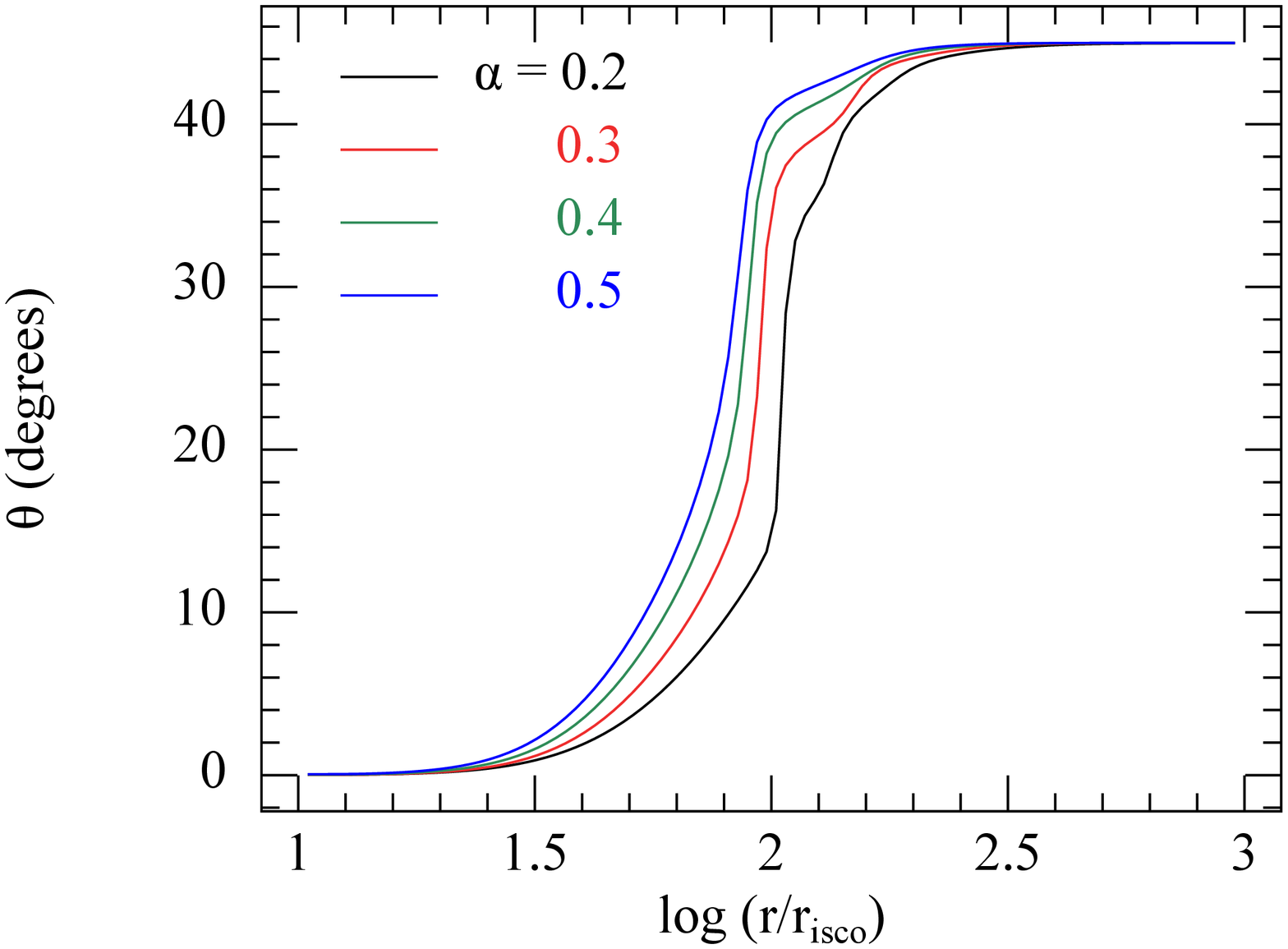}}
    \subfigure{\includegraphics[angle=0,width=0.33\textwidth]
       {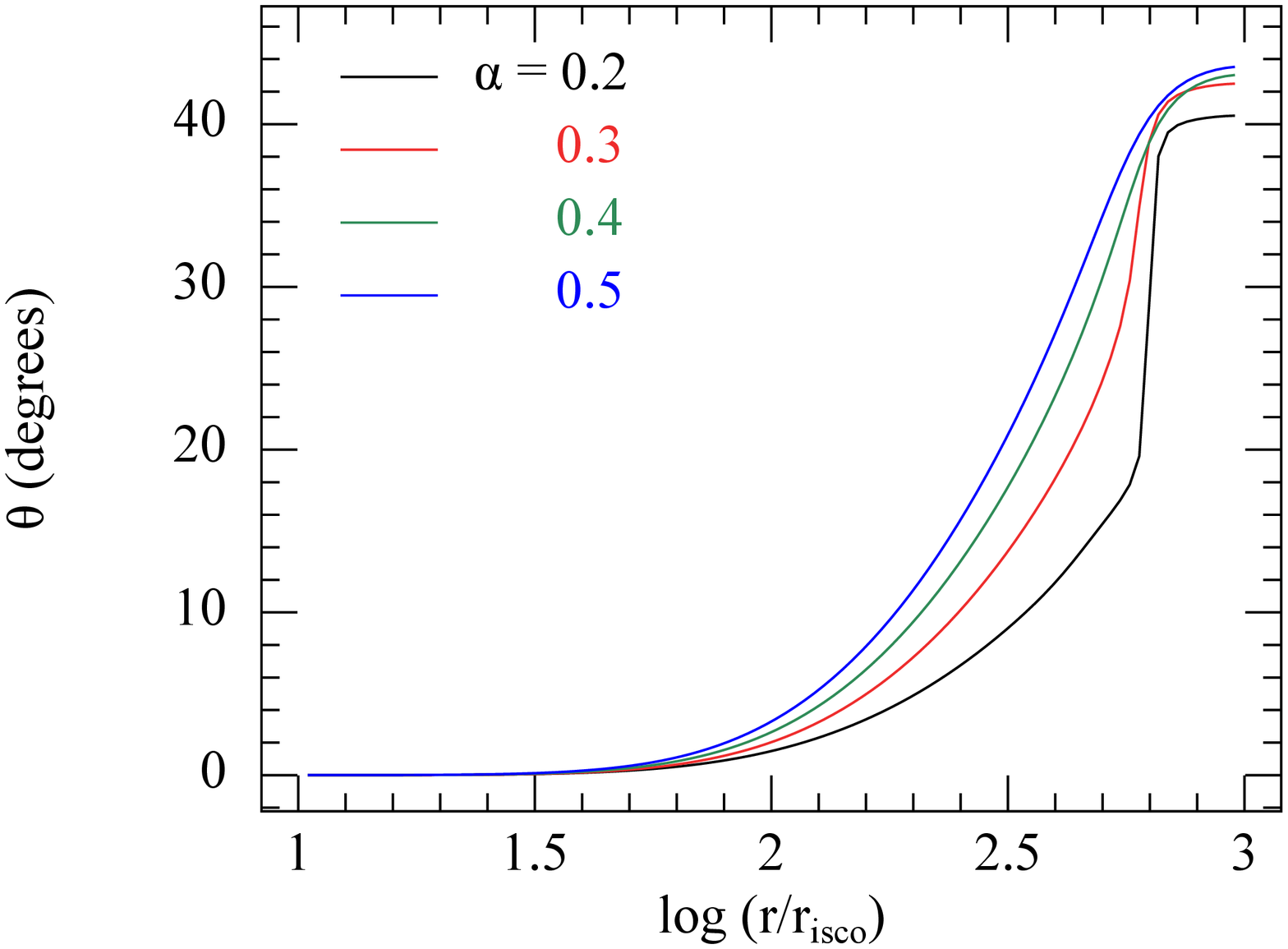}}
    \subfigure{\includegraphics[angle=0,width=0.33\textwidth]
       {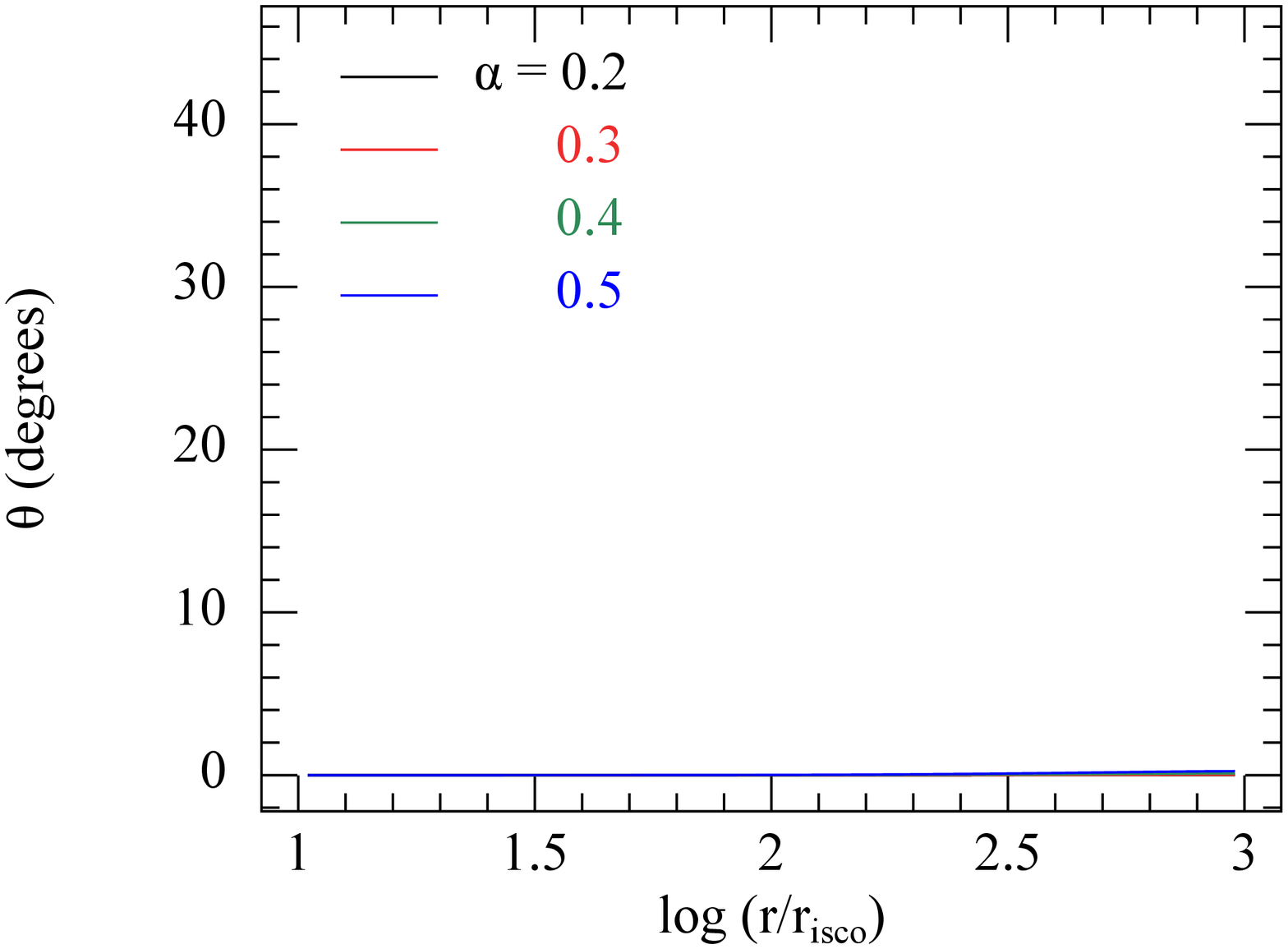}}
    \caption{Disc structures for the full effective viscosity simulations as
      in Fig.~\ref{og10}, but this time for initial misalignment $\theta =
      45^{\circ}$. From left to right the panels correspond to $t = 0.01$,
      $0.1$ and $1$ viscous times after the start of the calculation. When
      compared to \fref{const45} this figure shows that the nonlinear effects
      can significantly influence the evolution of the disc. Here the profile
      is noticeably steepened at $t=0.01t_{\rm visc}$ for all $\alpha$. By
      $t=0.1t_{\rm visc}$ the disc break is only maintained for small
      $\alpha$. For $\alpha = 0.2$ the break propagates outwards in the disc
      until reaching the outer edge of the grid (at $t = 0.13t_{\rm visc}$) at
      which point the disc aligns. For larger $\alpha$ the disc is smoothed
      before it aligns.}
    \label{og45}
\end{figure*}

In Figures~\ref{const45} \& \ref{og45} we show the disc structures for
simulations with $\theta_0=45^{\circ}$. There is now a noticeable difference
between the constant effective viscosity and full effective viscosity
simulations. Even for large $\alpha$ the disc profile is steepened and for
$\alpha \sim 0.2$ the disc breaks into two distinct planes. This break
persists for more than a tenth of the viscous timescale for the disc. The disc
aligns when the break propagates out to the edge of our grid which occurs at
$t = 0.13t_{\rm visc}$.

\subsubsection{$\theta = 60^{\circ}$}
\label{theta60}
\begin{figure*}
  \centering 
    \subfigure{\includegraphics[angle=0,width=0.33\textwidth]
       {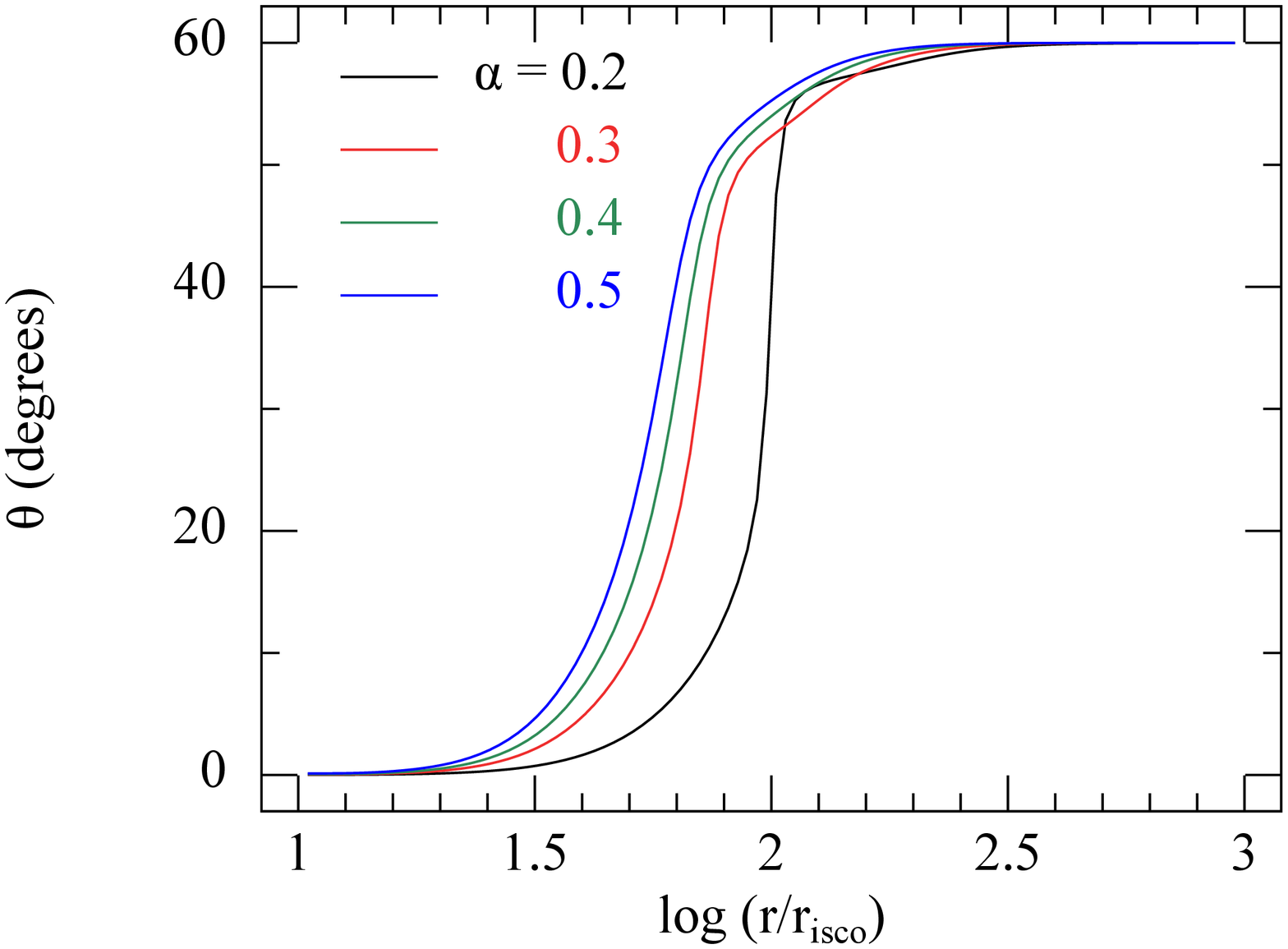}}
    \subfigure{\includegraphics[angle=0,width=0.33\textwidth]
       {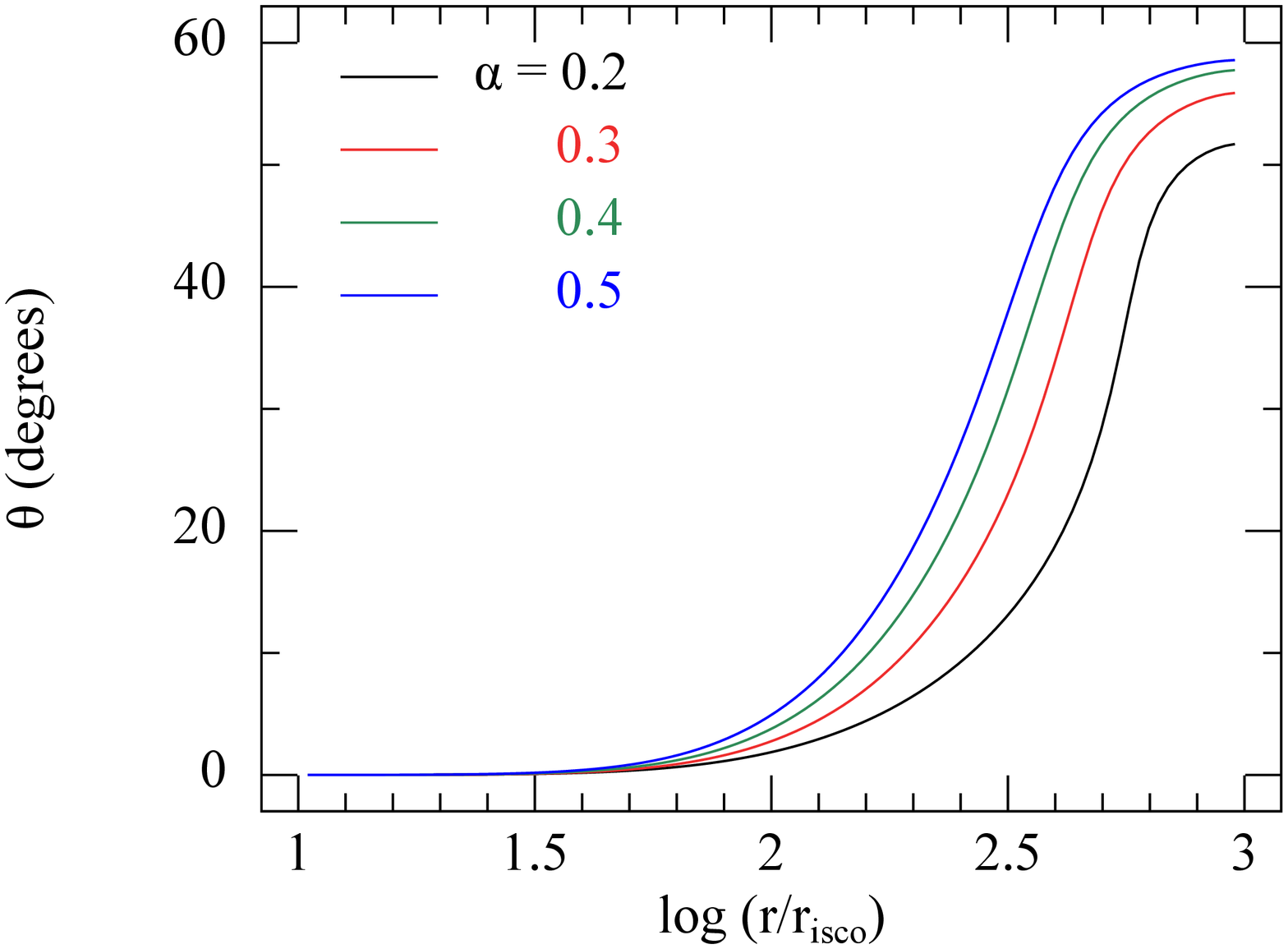}}
    \subfigure{\includegraphics[angle=0,width=0.33\textwidth]
       {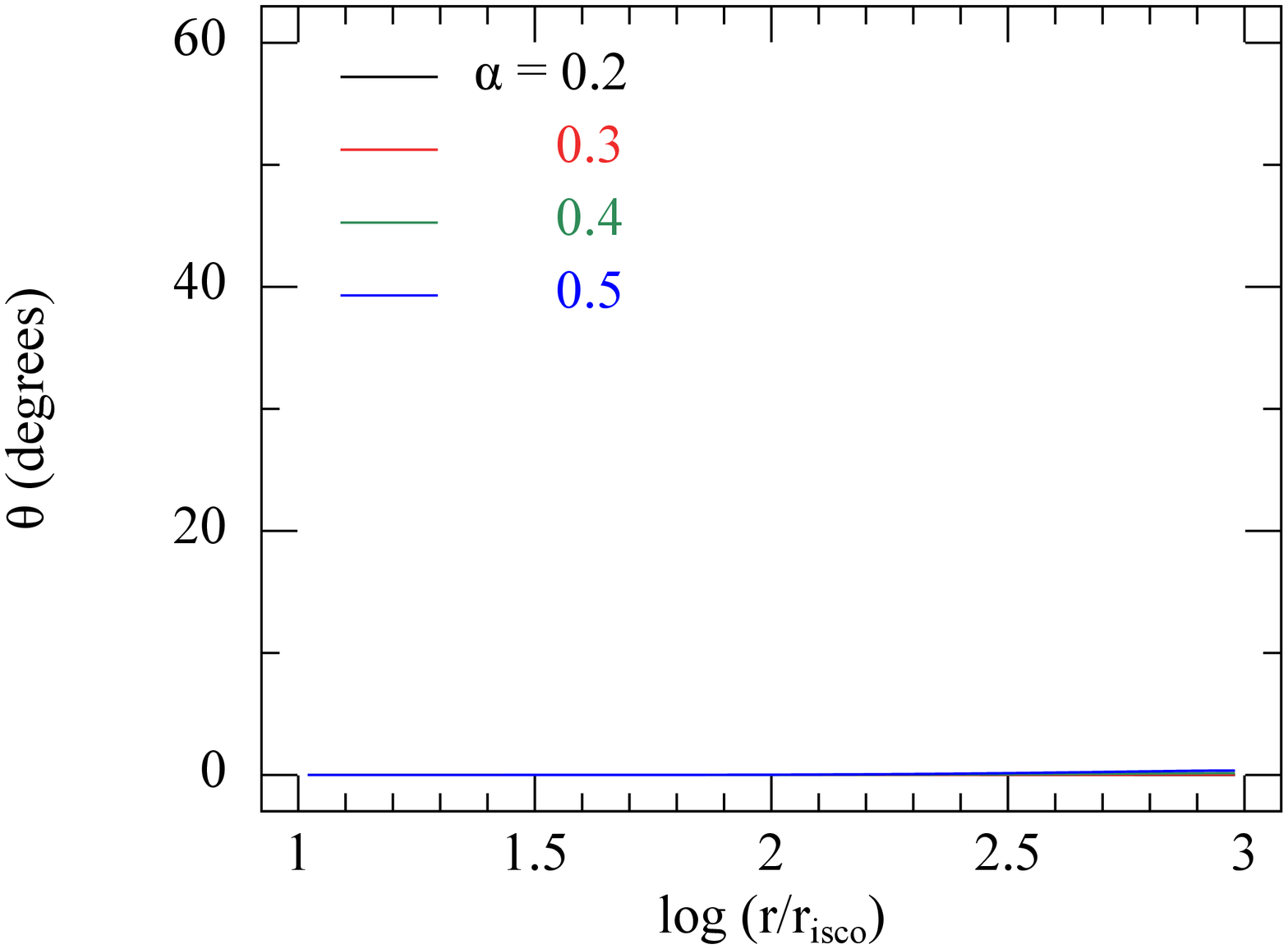}}
    \caption{Disc structures for the constant effective viscosity simulations
      as in Fig.~\ref{const10}, but this time for initial misalignment $\theta
      = 60^{\circ}$. From left to right the panels correspond to $t = 0.01$,
      $0.1$ and $1$ viscous times after the start of the calculation.}
    \label{const60}
\end{figure*}

\begin{figure*}
  \centering
    \subfigure{\includegraphics[angle=0,width=0.33\textwidth]
       {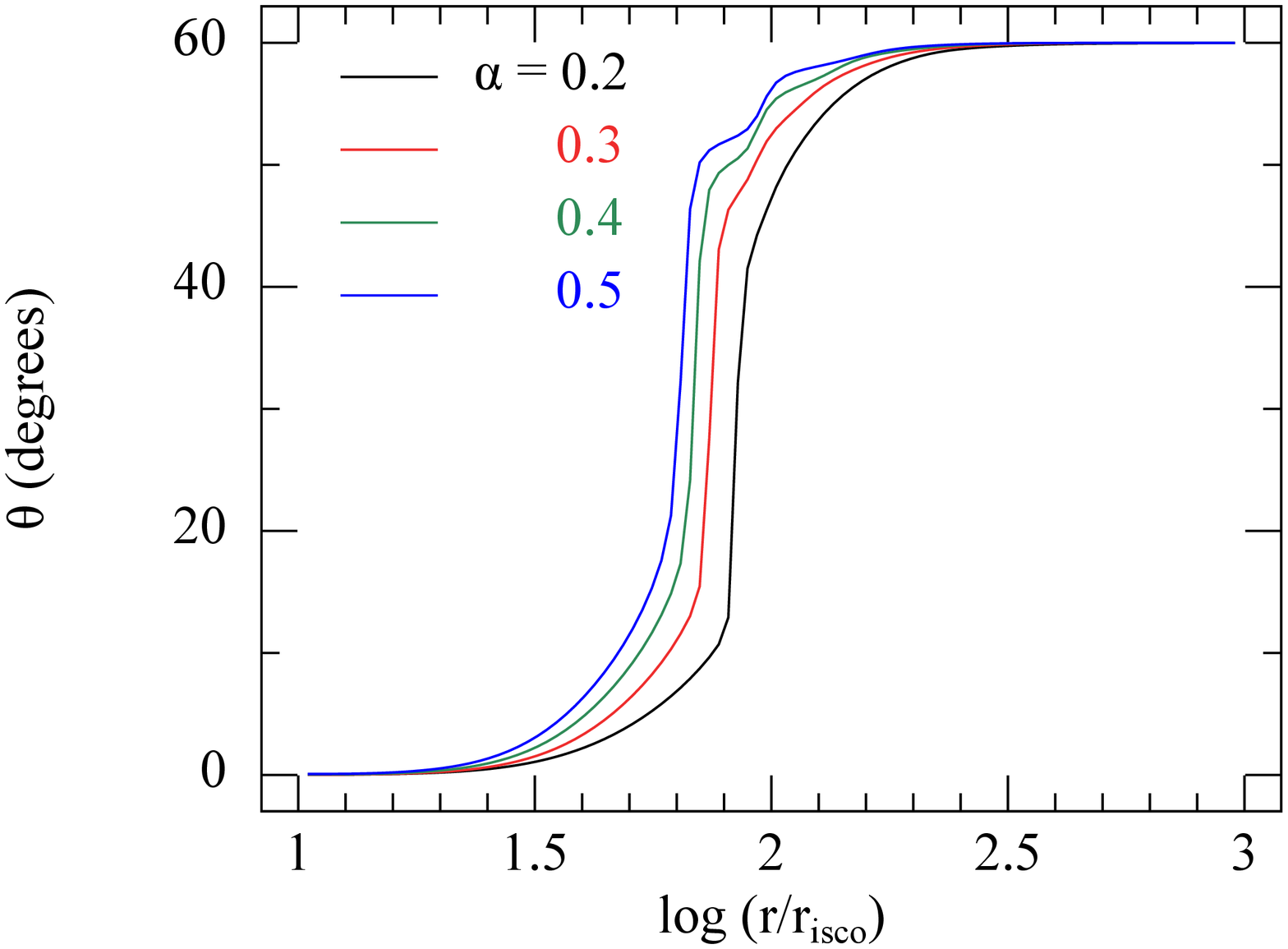}}
    \subfigure{\includegraphics[angle=0,width=0.33\textwidth]
       {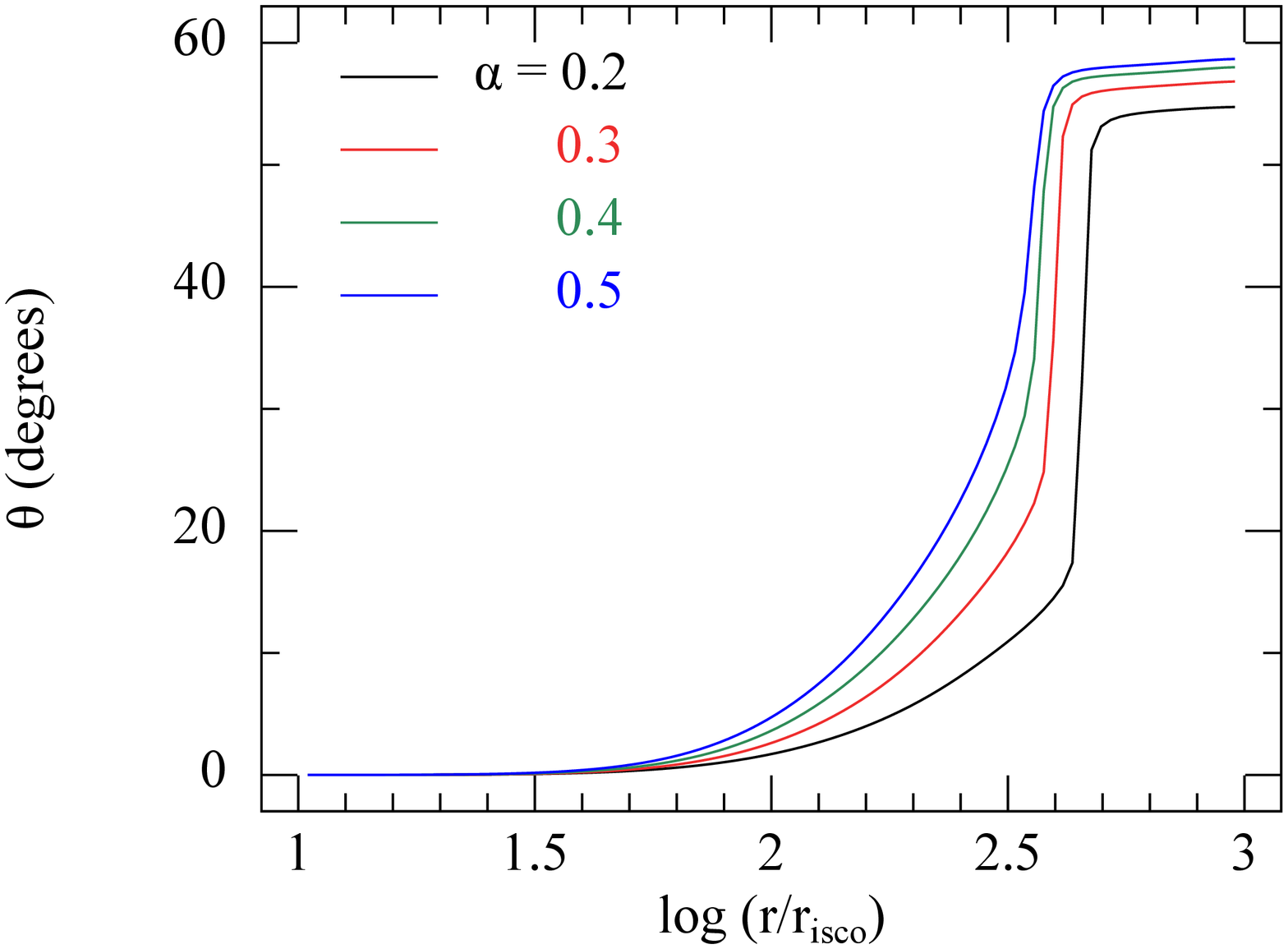}}
    \subfigure{\includegraphics[angle=0,width=0.33\textwidth]
       {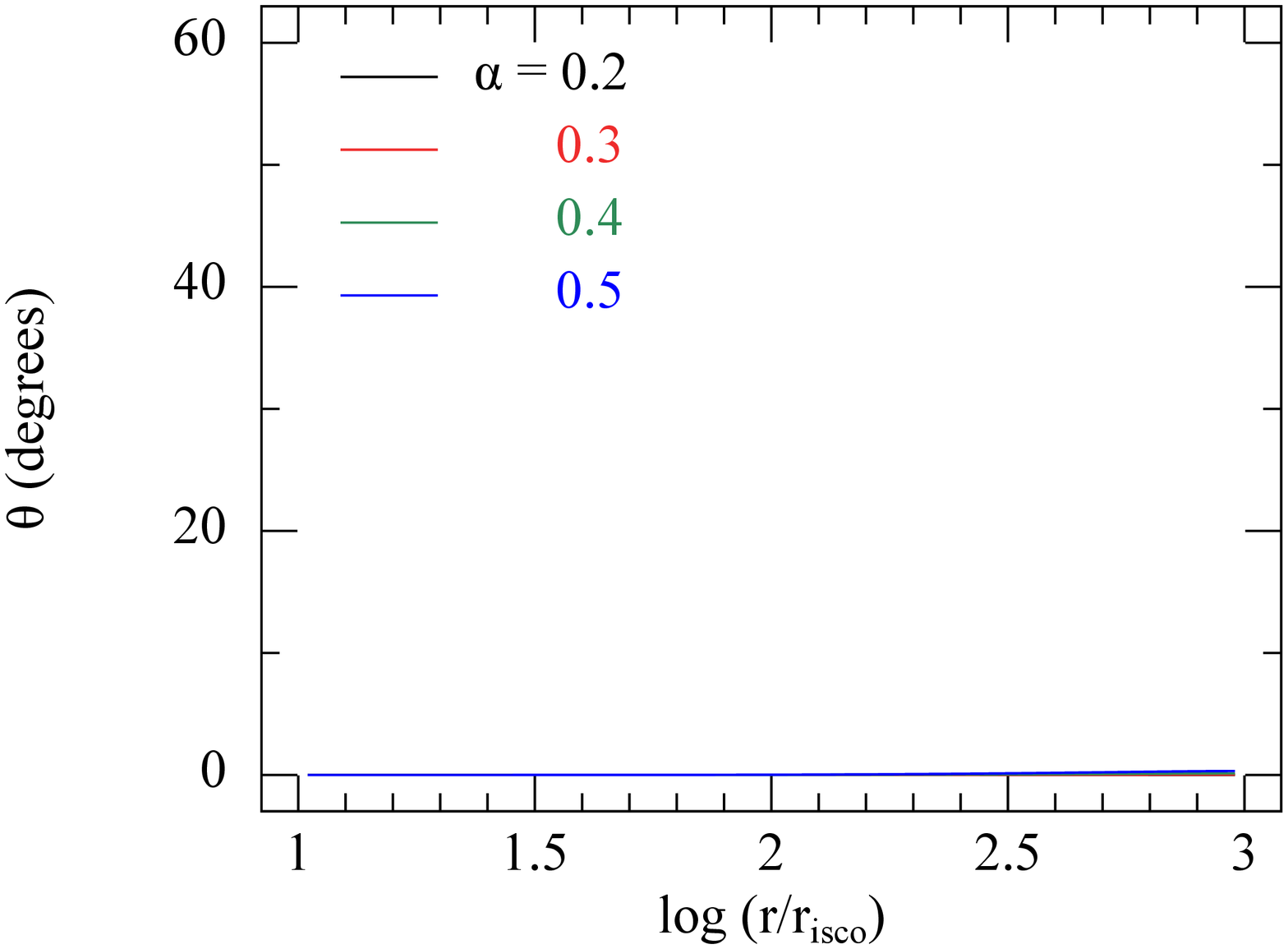}}
    \caption{Disc structures for the full effective viscosity simulations as
      in Fig.~\ref{og10}, but this time for initial misalignment $\theta =
      60^{\circ}$. From left to right the panels correspond to $t = 0.01$,
      $0.1$ and $1$ viscous times after the start of the calculation. When
      compared to \fref{const60} this figure shows that the nonlinear effects
      can significantly influence the evolution of the disc. Initially the
      disc profile is significantly steepened for all $\alpha$ and the disc
      breaks for all $\alpha$. This break propagates outwards in the disc
      until reaching the outer edge of the grid at which point the disc
      aligns.}
    \label{og60}
\end{figure*}

In Figures~\ref{const60} \& \ref{og60} we show the disc structures for
simulations with $\theta_0=60^{\circ}$. There is again a noticeable difference
between the constant effective viscosity and full effective viscosity
simulations. At $t=0.01t_{\rm visc}$ the disc profile is steepened noticeably
more than in the $\theta=45^{\circ}$ case and by $t=0.1t_{\rm visc}$ all of
the simulations have maintained a break. We can also see that at
$t=0.01t_{\rm visc}$ in the constant effective viscosity simulation with
$\alpha = 0.2$ it is possible for the disc to break without the nonlinear
effects.

These simulations suggests two main results of using the full effective
viscosities. First, discs can break for higher values of $\alpha$, and second,
discs can break for lower inclination angles.

There are two features of the simulations which we have not discussed so
far. The first is the wiggles in the disc at the top of the warp in Figures
\ref{og45} \& \ref{og60}. In these simulations the disc appears to try to
break into more than just two distinct planes. It may be that for smaller
$\alpha$ or larger $\left|\psi\right|$ the disc can break into more than two
planes. However we deem this beyond the scope of the present investigation.

Second, the torques on the disc in the strong warp regions generate rapid mass
flow rates into the ring inside the warp. This results in prominent spikes in
$\Sigma\left(R\right)$ inside regions with large $\left|\psi\right|$.
Increasing resolution only resolves these spikes better.  We therefore leave
the investigation of this effect to future work, which will require a
different numerical method.
%%%%%%%%%%%%%%%%%%%%%%%%%%%%%%%%%%%%%%%%%%%%%%%%%%%%%%%%%%%%%%%%%%%%%%%%%%%%%%
%% Discussion                                                               %%
%%%%%%%%%%%%%%%%%%%%%%%%%%%%%%%%%%%%%%%%%%%%%%%%%%%%%%%%%%%%%%%%%%%%%%%%%%%%%%
\section{Discussion}
\label{discussion}

We have simulated the evolution of a misaligned accretion event on to
a spinning black hole. Many previous simulations of this type assumed
that the effective viscosities governing the angular momentum
transport were constant. Here we have used effective viscosities
constrained to be consistent with the internal fluid dynamics of the
disc. We compare these results with those of equivalent simulations
which assume constant effective viscosities.  In the latter case we
find that for small enough $\alpha$ the precession induced by the LT
effect can cause the disc to break into two distinct planes if the
misalignment is large enough. We also find that if we allow nonlinear
effects \citep{Ogilvie1999} the disc can break for higher values of
$\alpha$ and smaller inclination angles. This suggests that nonlinear
effects significantly change the evolution of strongly warped
discs. The main effect is in reducing the magnitude of the effective
viscosities in the warped region, noticeably reducing $\nu_1$ more than
$\nu_2$. This reduces the disc's ability to smooth out the warp
induced by the LT torque and so prevent a break.  This is essentially
a modified form of the Bardeen--Petterson effect, where the warp is
steepened by the weakened disc response.

For small values of $\alpha$, $\nu_2 \gg \nu_1$.  Then the effective
viscosity that tries to flatten tilted rings of gas is much stronger than the
effective viscosity responsible for transporting angular momentum
radially. However the $\nu_1$ and $\nu_2$ torques responsible for
communicating angular momentum are proportional to $\boldsymbol{\ell}$ and
${\partial}\boldsymbol{\ell}/{\partial}R$ respectively. So to break the disc
we also need a large warp. As $\left|\psi\right|$ increases the $\nu_2$ torque
becomes comparatively stronger that the $\nu_1$ torque. When the discrepancy
between the magnitudes of the two torques is large enough the disc is
effectively unable to communicate radially. As the differential precession
induced by the LT effect acts faster at smaller radii, this allows the
vertical effective viscosity ($\nu_2$) to flatten each ring in turn into the
plane of the spinning black hole. To maintain a smooth (small amplitude) warp
the disc must be able to communicate radially on a timescale comparable to the
flattening of rings into the plane of the black hole. When this does not hold
the disc breaks.

For $\alpha \sim 1$ the disc can efficiently maintain a smooth warp. However
for $\alpha \sim 0.2-0.4$ and a large misalignment we show there is a strong
tendency for the disc to break under an external torque such as the LT
torque. We see no reason for this behaviour to change when $\alpha \le
0.1$. However there may be other nonlinear effects which become important
here. The next step is a full 3D hydrodynamical numerical investigation to
determine the disc response in this regime (e.g. \citealt{NP2000};
\citealt{LP2010}).

When the disc breaks it splits into a co-- or counter--aligned inner disc and
a misaligned outer disc. This break can propagate outwards and (dependent on
$\alpha$ and $\theta_0$) may persist until it reaches the outer regions of the
disc. At this point the whole disc is co-- or counter--aligned with the hole
spin. This suggests that a break in the disc can be a long--lived feature,
lasting for roughly the alignment timescale. We note that although we have not
included any simulations of counter--aligning discs, it is already known that
for $J_{\rm d}/J_{\rm h} \ll 1$ there is a symmetry in the behaviour of discs
of this type around $\theta=90^{\circ}$ (\citealt{Kingetal2005} and
\citealt{LP2006}). Therefore the same behaviour would be seen in discs that
are counter--aligning with the black hole (i.e. a misalignment of
$120^{\circ}$ would produce the same dynamics with a counter--aligning disc as
the $60^{\circ}$ case produces here).

For observational purposes it is important to compare the alignment timescale
$t_{\rm al}$ with the lifetime $t_{\rm life}$ of the system. If $t_{\rm al} >
t_{\rm life}$, the centre of the disc is never aligned with its warped outer
plane. In an accreting stellar--mass black--hole binary this would mean that
jets (which propagate along the axis of the central parts of the disc, and so
the spin of the black hole) are in general misaligned with respect to the
orbital plane. This is seen in the black--hole binary GRO J1655-40
(\citealt{Martinetal2008} and references therein). Similarly the
continuum--fitting method for estimating black--hole spins
(\citealt{Kulkarnietal2011} and references therein) uses the assumption that
the black hole spin is aligned with the orbital axis in estimating the
radiating disc area and hence the ISCO radius. It therefore implicitly
requires $t_{\rm al} < t_{\rm life}$. In our simulations it is clear that
discs are able to remain warped or broken for significant fractions of their
viscous timescale. However we note that for strong warps there may be other
nonlinear effects not included in the analysis of \citet{Ogilvie1999} or
additional dissipation caused by fluid instabilities (cf
\citealt*{Gammieetal2000}). These may impact upon the alignment timescale.

It is also possible that the disc is never able to align fully; for example an
accretion disc in a misaligned black hole binary system may be twisted one way
by the LT effect and a different way by the binary torque
\citep[e.g.][]{Martinetal2009}. We therefore suggest careful consideration
before assuming that black hole accretion discs are aligned -- particularly as
\citet{LP2006} suggest that the accretion rate through warped discs is
significantly enhanced, which suggests the possibility that strongly accreting
discs are also strongly warped.

Disc breaking can have some important observational consequences. For example
accretion discs in binary systems often have an outer disc aligned to the
binary plane. However the inner disc (and so the direction of any jets) must
be aligned with the black hole spin, which may be significantly misaligned
with respect to the binary axis. Similarly breaks in protoplanetary discs may
occur through the internal disc response to warping during formation of the
disc, rather than being driven by the presence of a planet.

In this paper we have used a 1D Eulerian ring code to calculate the
evolution of misaligned accretion events on to a spinning black
hole. We caution that there is still much work to do before we can
fully understand the evolution of such a disc. For example our
treatment requires gas to orbit on circular rings, inherently
excluding any effects which might break the cylindrical symmetry in
the local disc plane. We also assume that the disc responds
viscously, with no wave--like behaviour. The method also prevents us
from examining exactly what happens in the disc break: by its nature,
it occurs over only over a small number of rings and is therefore
poorly resolved -- however this does not affect our conclusion that
the disc can break.  The most frustrating restriction is the
difficulty in simulating small enough $\alpha$: the simulations we
present here suggest that for small $\alpha$ nonlinear fluid effects
can be particularly important. We will address these restrictions in
future investigations.

We remark finally that although we have considered alignment effects
induced by the LT effect on an accretion disc around a spinning black
hole, the torque between a misaligned binary and an external accretion
disc has a very similar form \citep*{Nixonetal2011b}. Thus we expect
similar phenomena to appear in that case. The enhanced accretion rates
through the disc due to the dissipation in the warp \citep{LP2006}
would increase the mass flow rate on to the binary. Therefore the
coalescence timescale for the binary would be reduced further
\citep{Nixonetal2011a}.

%%%%%%%%%%%%%%%%%%%%%%%%%%%%%%%%%%%%%%%%%%%%%%%%%%%%%%%%%%%%%%%%%%%%%%%%%%%%%%
%% Acknowledgements                                                         %%
%%%%%%%%%%%%%%%%%%%%%%%%%%%%%%%%%%%%%%%%%%%%%%%%%%%%%%%%%%%%%%%%%%%%%%%%%%%%%%
\section*{Acknowledgments}
\label{acknowledgements}
We thank the anonymous referee for valuable comments which helped to improve
the manuscript. We thank Gordon Ogilvie for providing us with the routine to
calculate the nonlinear viscosity coefficients. We also thank Jim Pringle,
Gordon Ogilvie, Richard Alexander and Giuseppe Lodato for useful
discussions. CJN holds an STFC postgraduate studentship. We acknowledge the
use of \textsc{splash} \citep{Price2007} for generating the figures. Research
in theoretical astrophysics at Leicester is supported by an STFC Rolling
Grant. This research used the ALICE High Performance Computing Facility at the
University of Leicester.  Some resources on ALICE form part of the DiRAC
Facility jointly funded by STFC and the Large Facilities Capital Fund of BIS.
%%%%%%%%%%%%%%%%%%%%%%%%%%%%%%%%%%%%%%%%%%%%%%%%%%%%%%%%%%%%%%%%%%%%%%%%%%%%%%
%% Bibliography                                                             %%
%%%%%%%%%%%%%%%%%%%%%%%%%%%%%%%%%%%%%%%%%%%%%%%%%%%%%%%%%%%%%%%%%%%%%%%%%%%%%%
\bibliographystyle{mn2e} 
\bibliography{nixon}
\end{document}